\documentclass[12pt]{article}

\usepackage{newtxtext,newtxmath}

\usepackage{graphicx}
\usepackage{color}
\usepackage{xcolor}

\usepackage[letterpaper,margin=1in]{geometry}

\renewenvironment{abstract}
	{\quotation}
	{\endquotation}

\date{}

\makeatletter
\renewcommand{\fnum@figure}{\textbf{Figure \thefigure}}
\renewcommand{\fnum@table}{\textbf{Table \thetable}}
\makeatother

\usepackage{scicite}

\usepackage{url}
\usepackage{hyperref}

\newcommand\farcs{\mbox{$.\!\!^{\prime\prime}$}}%

\newcommand{\correct}[1]{\textcolor{black}{ #1}}
\newcommand{\correctn}[1]{\textcolor{black}{ #1}}

\def\Hb{H$\beta$}

\def\Cii{[C\,{\sc ii}]}
\def\Ciium{[C\,{\sc ii}]\,$158\,{\rm \mu m}$}

\def\Oiiilong{[O\,{\sc iii}]\,$5008\,{\rm \AA}$}

\def\Oiiiboth{[O\,{\sc iii}]\,$\lambda\lambda4960,\,5008\,{\rm \AA}$}
\def\Hb{H$\beta$}

\def\ltsima{$\buildrel<\over\sim$}
\def\la{\lower.5ex\hbox{\ltsima}~}
\def\gtsima{$\buildrel>\over\sim$}
\def\ga{\lower.5ex\hbox{\gtsima}~}
\def\deg~{$^{\circ}$}
\def\quintet{A2744-{\em Quintet}}

\def\scititle{
    Early massive galaxy formation in the core of a galaxy protocluster 650 million years after the Big Bang
}
\title{\bfseries \boldmath \scititle}

\author{
	Yoshinobu~Fudamoto$^{1\ast}$,
        Yurina Nakazato$^{2}$,
        Daniel Ceverino$^{3,4}$,
        Luis Colina$^{5}$,\and
        Takuya Hashimoto$^{6,7}$,
        Akio K. Inoue$^{8,9}$,
        Yoichi Tamura$^{10}$,
        Naoki Yoshida$^{2,11,12}$,\and
        Yongda Zhu$^{13}$,
        Yuma Sugahara$^{8,9}$,
        Santiago Arribas$^5$,
        Javier {\'A}rvarez-M{\'a}rquez$^5$,\and
        Tom Bakx$^{14}$,
        Carmen Blanco Prieto$^5$,
        Luca Costantin$^5$,
        Alejandro Crespo G{\'o}mez$^{15}$,\and
        Masato Hagimoto$^{10}$,
        Takeshi Hashigaya$^{16}$,
        Hiroshi Matsuo$^{17,18}$,
        Rui Marques-Chaves$^{19}$,\and
        Ken Mawatari$^{8,9}$,
        Ikki Mitsuhashi$^{20}$,
        Wataru Osone$^{6}$,
        Miguel Pereira-Santaella$^{21}$,\and
        Hideki Umehata$^{5}$,
        Callum Witten$^{19}$,
        Yi W. Ren$^{9}$,\and
	\small$^{1}$Center for Frontier Science, Chiba University, 1-33 Yayoi-cho, Inage-ku, Chiba 263-8522, Japan\and
    \small$^2$Department of Physics, The University of Tokyo, 7-3-1 Hongo, Bunkyo, Tokyo 113-0033, Japan\and
    \small$^3$Universidad Autonoma de Madrid, Ciudad Universitaria de Cantoblanco, E-28049 Madrid, Spain\and
    \small$^4$CIAFF, Facultad de Ciencias, Universidad Autonoma de Madrid, E-28049 Madrid, Spain\and
    \small$^5$Centro de Astrobiolog{\'i}a (CAB), CSIC-INTA, Ctra. de Ajalvir km 4, Torrej{\'o}n de Ardoz, E-28850, Madrid, Spain\and
    \small$^6$Division of Physics, Faculty of Pure and Applied Sciences, University of Tsukuba, Tsukuba, Ibaraki 305-8571, Japan\and
    \small$^7$Tomonaga Center for the History of the Universe (TCHoU), Faculty of Pure and Applied Sciences, University of\and\small Tsukuba, Tsukuba,
Ibaraki 305-8571, Japan\and
    \small$^8$Waseda Research Institute for Science and Engineering, Faculty of Science and Engineering, Waseda University,\and\small 3-4-1 Okubo, Shinjuku, Tokyo 169-8555, Japan\and
    \small$^9$Department of Pure and Applied Physics, School of Advanced Science and Engineering, Faculty of Science\and\small and Engineering, Waseda University, 3-4-1 Okubo, Shinjuku, Tokyo 169-8555, Japan\and
    \small$^{10}$Department of Physics, Graduate School of Science, Nagoya University, Nagoya 464-8602, Japan\and
    \small$^{11}$Kavli Institute for the Physics and Mathematics of the Universe (WPI), UT Institute for Advanced Study,\and\small The University of Tokyo, Kashiwa, Chiba 277-8583, Japan\and
    \small$^{12}$Research Center for the Early Universe, School of Science, The University of Tokyo, 7-3-1 Hongo, Bunkyo,\and\small Tokyo 113-0033, Japan\and
    \small$^{13}$Steward Observatory, University of Arizona, 933 North Cherry Avenue, Tucson, AZ 85721, USA\and
    \small$^{14}$Department of Space, Earth, \& Environment, Chalmers University of Technology, Chalmersplatsen,\and\small SE-4 412 96 Gothenburg, Sweden\and
    \small$^{15}$Space Telescope Science Institute (STScI), 3700 San Martin Drive, Baltimore, MD 21218, USA\and
    \small$^{16}$Department of Astronomy, Kyoto University Sakyo-ku, Kyoto 606-8502, Japan\and
    \small$^{17}$National Astronomical Observatory of Japan, 2-21-1 Osawa, Mitaka, Tokyo 181-8588, Japan\and
    \small$^{18}$Graduate University for Advanced Studies (SOKENDAI), 2-21-1 Osawa, Mitaka, Tokyo 181-8588, Japan\and
    \small$^{19}$Geneva Observatory, Department of Astronomy, University of Geneva, Chemin Pegasi 51, CH-1290 Versoix,\and\small Switzerland\and
    \small$^{20}$Department for Astrophysical \& Planetary Science, University of Colorado, Boulder, CO 80309, USA\and
    \small$^{21}$Instituto de F{\'i}sica Fundamental (IFF), CSIC, Serrano 123, E-28006, Madrid, Spain\and
    \small$^\ast$Corresponding author. Email: yoshinobu.fudamoto@gmail.com\and
}

\begin{document} 

\maketitle

\begin{abstract} \bfseries \boldmath
\correct{Rest-frame optical observations with the James Webb Space Telescope (JWST) have uncovered a population of massive galaxies, exceeding $10^{10}$ solar masses, present less than a billion years after the Big Bang.
The large stellar masses of these galaxies require an efficient
conversion of baryons into stars, which may exceed theoretical expectations.
However, the formation process of massive galaxies so early in the Universe’s history is perplexing, as observations provide limited information to constrain their evolutionary pathways.
Here, we present multi-wavelength observations of a galaxy complex consisting of at least five galaxies within a $\sim10\,{\rm kpc}$ region,
referred to as the \quintet, using JWST and the Atacama Large Millimeter/submillimeter Array.
This system, located in the core of a galaxy protocluster at approximately 650 million years after the Big Bang, reveals the detailed physical processes involved in the formation of massive galaxies. These processes include a dynamic cycles of merger induced gas stripping, leading the temporal termination of star formation, and recycling of the stripped gas, with subsequent enhancement of star formation in other galaxies of the system, \correctn{which is expected to evolve into massive galaxies that host more than $10^{10}$ solar masses of stars.}
The new observations represent the first comprehensive evidence of a massive galaxy formation through gas-rich, multiple-galaxy mergers induced by a dense protocluster environment in the $650\,{\rm Myrs}$ after the Big Bang. 
Our results suggest that the protocluster core is indeed one of the main drivers of efficient galaxy formation and rapid evolution in the early Universe, as predicted by theoretical studies.
}
\end{abstract}

\noindent
The James Webb Space Telescope (JWST) has been providing an unprecedented view of galaxy evolution in the early Universe by allowing access to faint limits of rest-frame ultraviolet (UV) to near-infrared emission from galaxies at redshift $z>7$ \cite{Carniani2024Natur,Finkelstein2024,Perez-Gonzalez2025,Naidu2025}. 
These observations reveal that some galaxies had already assembled substantial stellar masses \correct{, exceeding $10^{10}$ solar masses (${\rm M_{\odot}}$)} less than a billion years after the Big Bang, implying that star formation in the early galaxies was rapid and highly efficient \cite{Carnall2023Natur,Xiao2024,deGraaff2025,Weibel2025}.
\correct{The derived baryon conversion efficiencies for these massive galaxies are often at, or exceed, the upper limits from theoretical predictions\cite{Dekel2023,Ferrara2023,Ceverino2024}, requiring high gas densities, possibly driven by violent processes such as galaxy-galaxy mergers.}
On the other hand, spectral energy distribution (SED) modeling of high redshift galaxies indicates a diversity of star-formation histories among these early systems from intense, short-lived bursts \cite{Tacchella2023}, to recent or temporal quenching of star formation \cite{Looser2024, Weibel2025}, and even galaxies that have rejuvenated star formation following their quiescent phases \cite{Witten2025}.
\correct{Actual drivers of the rapid galaxy growths as well as diverse transitions of star formation activities are not observationally constrained.}

\begin{figure}[h]
    \centering
    \includegraphics[width=\linewidth]{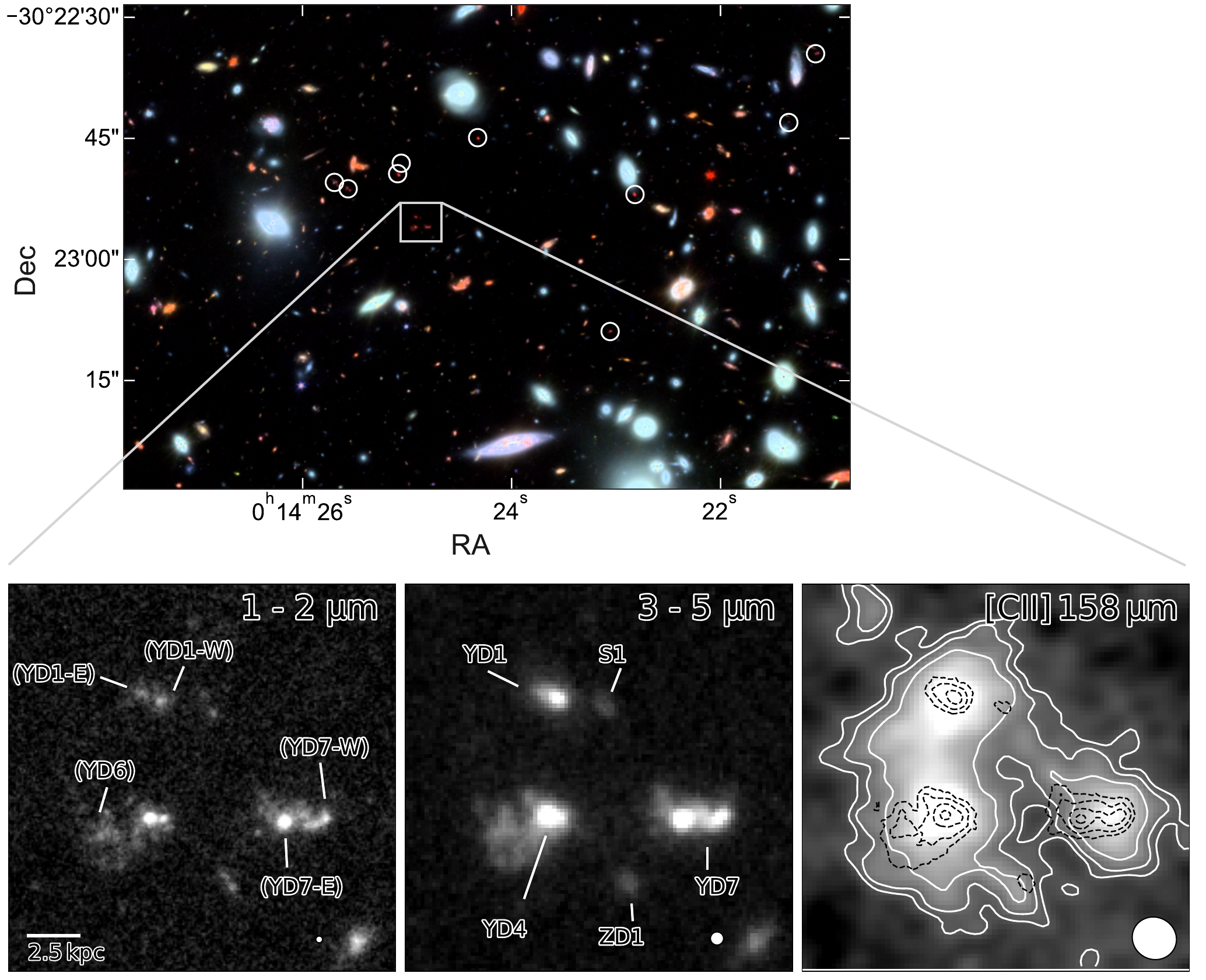}
    \caption{{\bf JWST images of the \quintet\ at $\mathbf{z=7.9}$}:
	Upper image shows a false color image of a $60^{\prime\prime}\times90^{\prime\prime}$
	region around the \quintet\ using F444W (red), F277W (green), and F150W images (blue). \correctn{White circles indicate galaxies belong to A2744-z7p9OD, a galaxy protocluster at $z=7.9$ \cite{Witten2025-2}. The white square indicate the \quintet.} 
    \correct{Lower left and middle panels show the stacked images of the \quintet\ obtained by JWST's short and long wavelength filters, respectively.}
	Lower right panel shows the ALMA \Ciium\ emission line map. 
    The white contours show the $2,3,5\,\sigma$ levels of the \Ciium\ emission line, and black dashed contours show $6,12,24,48\,\sigma$ signal in the stacked \correctn{JWST's} long wavelength image.
	Labels show names of each member galaxies of the \quintet\ and labels in parentheses show names of sub-components.
    The white bar in the lower left corner shows physical $2.5\,{\rm kpc}$ scale \correct{in the image plane without lensing corrections. The white ellipses in the lower \correctn{right corners in each panel} show full width at half maximums of the point spread functions and the synthesized beam size of JWST and ALMA observations, respectively.}}
    \label{fig:quintet}
\end{figure}

Environmental effects, including galaxy--galaxy mergers, are thought to play \correct{one of the key roles} in transforming star-formation activity at least in the later time of the Universe \cite{Patton2020,Garduno2021,Ellison2022,Puskas2025}, leading to accelerated galaxy evolution \correct{which results} in both active star-formation and quenching through dynamical processes.
\correct{To investigate how galaxy environments drive galaxy evolution at very early times, the study of galaxy over-densities and protoclusters at the epoch of reionization is particularly relevant \cite{Helton2024,Arribas2024}.}
 In this context, we conduct a multi-wavelength study of \correct{galaxy evolution processes within one of the densest galaxy environments known so far at redshift of $z>7$}, the \quintet\ (Figure \ref{fig:quintet}).
This system is located within the galaxy protocluster A2744-z7p9OD, which is one of the highest-redshift spectroscopically confirmed galaxy protoclusters known to date \cite{Morishita2023,Hashimoto2023}.
None of the galaxies in the \quintet\ show evidence of \correct{an} active galactic nuclei (AGN), as inferred from their extended rest-frame \correct{UV} morphologies and the absence of broad Balmer lines or strong UV/optical emission lines seen from JWST observations (see Methods). 
This indicates that their observed emission is not dominated by black hole accretion but by their star-formation.
The \quintet\ thus offers a unique opportunity to probe the star formation processes governing early galaxy growth in a dense environment.

\begin{figure}[h] 
	\centering
	\includegraphics[width=1.0\textwidth]{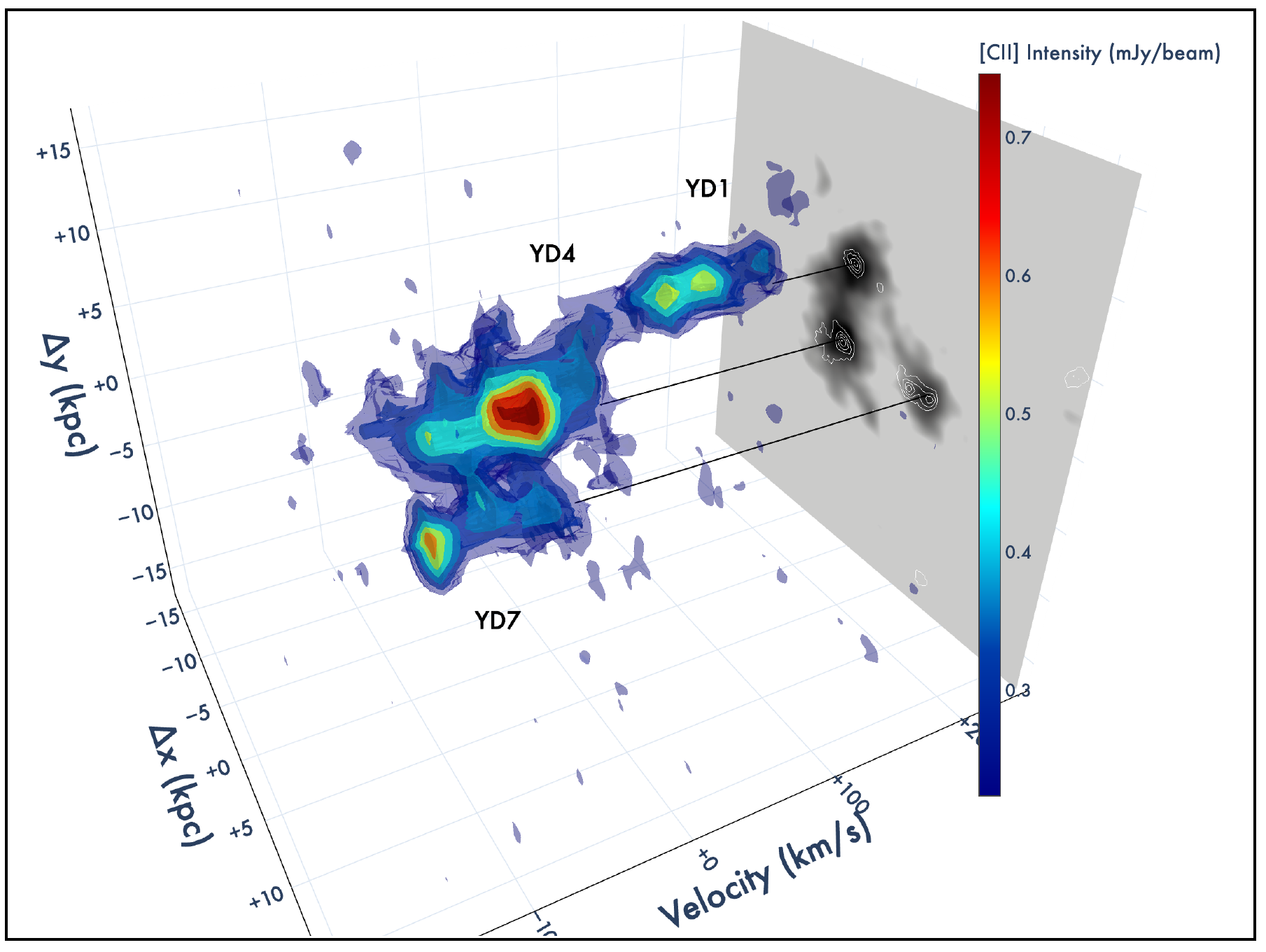} 
	\caption{\textbf{Observed \Ciium\ emission lines and the gas structure of the A2744-the Quintet.}
	3-D visualization of the \Ciium\ emission line obtained in our observations. 
	The bluest color starts from $3\,\sigma$ signal and reddest color ends at $10\,\sigma$ signal.
    \correct{The velocity reference is the strongest signal in the data, which corresponds to the \Cii\ emission from YD4.}
	The 2-D image on the back shows moment-0 image of the \Ciium\ emission line overlaying contours of a stacked image using JWST's long wavelength data ($6\sigma$ -- $48\,\sigma$).
	An interactive version of the 3-D visualized \Cii\ emission map is available as a html format file in \url{https://github.com/yfudamoto/data-the_Quintet}.
    }
	\label{fig:C2map} 
\end{figure}

\correct{Through Atacama Large Millimeter/submillimeter Array (ALMA) observations, we target, the singly-ionized carbon emission line, \Ciium\ and dust continuum of the \quintet. 
Both of these emission are known to trace star-forming, neutral interstellar medium that fuels star formation \cite{Zanella2018,Dessauges2020,Madden2020,Fudamoto2025}.
Multiple components of the \quintet\ are significantly detected in the \Ciium\ (Figure \ref{fig:quintet}, lower left panel) and dust continuum (Methods).}
\correct{In particular, as seen in the 3-D visualized map (Figure \ref{fig:C2map}), the \Cii\ emission extends \correctn{across the system's velocity range of} $\Delta {\rm d}v\sim300\,{\rm km/s}$ along the line of sight with highly clumpy morphology.
The three brightest \Cii\ clumps correspond spatially and spectrally to the three components that have rest-frame UV/optical detections seen in JWST (YD1, YD4 and YD7-W). These bright \Cii\ clumps have \Cii-based redshift consistent with those measured using \Oiiiboth\ lines \cite{Hashimoto2023}.
Thus, these observations show that these components represent gas-rich, star-forming clumps under the process of merging in the \quintet\ (Table \ref{tab:quintet_sfr}).}

In addition to \correct{bright \Cii-emitting clumps, diffuse} \Ciium\ emission is also detected in the circumgalactic regions.
\correctn{Unlike the bi-conical or symmetric, centrally diverging morphologies, without a preferred alignment along intergalactic axes, expected for star formation-driven outflows\cite{Walter2002}, the diffuse \Cii\ emission seen the \quintet\ is primarily concentrated along filamentary structures connecting each \Cii-emitting clumps.}
\correct{This spatial configuration indicates that the emission traces gas tidal bridges formed through on-going mergers, rather than gas expelled by feedback.
This interpretation is in line with the higher projected velocity dispersion of \Cii\ emission in the bridging structure (Figure \ref{fig:velocity} in Methods).
Finally, some of the galaxies and sub-components (ZD1, S1, and YD7-E) lack significant detections of \Ciium\ and dust continuum emission, indicating the absence of a substantial neutral gas reservoir (Methods).
The presence of gas tidal bridges and the lack of neutral gas reservoir in parts of the \quintet\ suggest the star-forming interstellar medium  has been at least partially removed through tidal stripping during the on-going mergers.
Furthermore, the presence of bright \Ciium\ emitting clumps connected by the filamentary tidal gas bridges suggest that gas stripped during the ongoing mergers is being channeled to these star-forming regions.
This implies that merger-driven tidal interactions are not only removing gas but also redistributing it into these \Cii-luminous clumps, fueling their star formation.
We further confirm these phenomena using JWST observations in the following.}

\begin{figure}[h]
    \centering
    \includegraphics[width=0.85\linewidth]{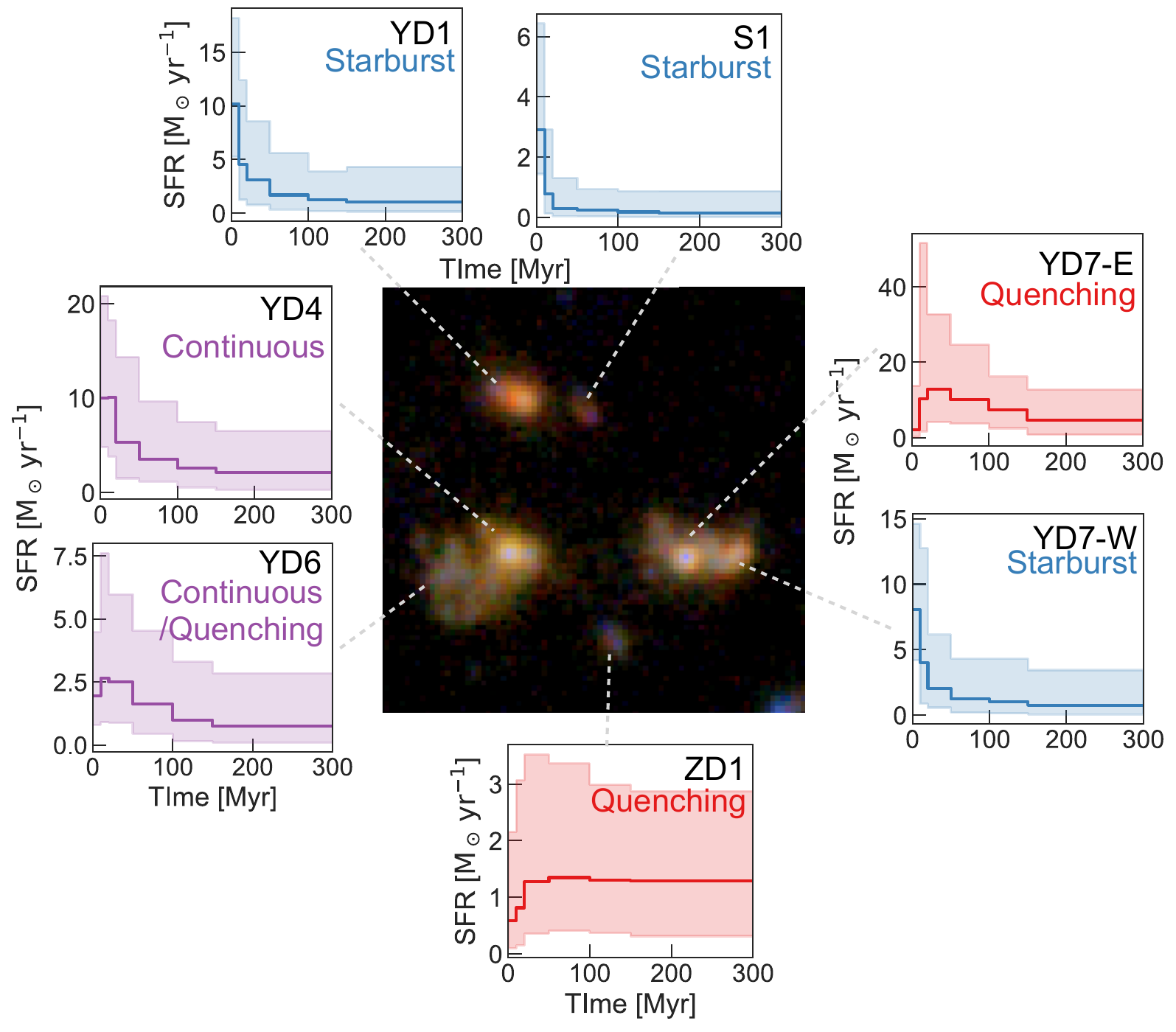}
    \caption{\textbf{Star formation histories of member galaxies in \quintet}:
    Middle panel shows a false color image of the \quintet\ using the stacked images of the observed wavelengths of $1.2$ -- $2.1\,{\rm \mu m}$ in red, $2.5$ -- $3.6\,{\rm \mu m}$ in green, and $4.1$ -- $4.8\,{\rm \mu m}$ in blue. 
    In other panels, spectral energy distributions (SEDs) modeling show distinct star formation histories (SFHs) of galaxies in the \quintet, showing the outcomes of gas transportation within the system \correct{(see Methods for SED fitting and the definition of star formation modes).}
    The recent quenching of of YD7-E and ZD1, \correct{the continuous star-formation of YD4 and YD6}, and the starbursts of YD1, S1, and YD7-W show the results of gas stripping, compression, and recycling as seen from the spatial	distribution of the \Ciium\ emission lines. 
    }
    \label{fig:SED}
\end{figure}

To constrain the physical properties of galaxies in the \quintet,
we performed SED modeling using JWST data to estimate their star-formation histories (Figure~\ref{fig:SED} and Table~\ref{tab:quintet_sfr}).
\correct{From the SED modeling, we find a diversity of star formation activity within the \quintet.}
YD4 shows a strong Balmer break with strong emission lines, indicating the existence of both \correct{matured and young} stellar populations and continuing intense star formation activity\cite{Witten2025-2}.
Both YD7-E and ZD1 show blue UV spectral slopes and prominent Balmer breaks with little or no detectable UV and optical emission lines \correctn{\cite{Hashimoto2023}} (see also Methods). 
These features of YD7-E and ZD1 are a characteristic of galaxies that have recently quenched their star-formation at least $\sim 10\,{\rm Myrs}$ ago \cite{Looser2024}. 
The derived star-formation histories support this interpretation, indicating a rapid decline in star formation activity, \correctn{as indicated by their SFR ratios over $10\,{\rm Myr}$ and $100\,{\rm Myr}$ time scales: $\rm{SFR_{10Myr}/SFR_{100\,Myr}} = 0.1^{+0.7}_{=0.1}$ and $0.3^{+0.9}_{-0.2}$ for YD7-E and ZD1, receptively (Table \ref{tab:quintet_sfr})}.
In contrast, YD1 and S1 show prominent optical emission lines, such as \Oiiiboth\ and \Hb, but only weak stellar continuum emission \correctn{\cite{Hashimoto2023}}. \correct{Thus, unlike ZD1 and YD7-E}, YD1 and S1 are undergoing a recent burst of star formation activity and may have formed recently from the metal-enriched gas in the surroundings of the system, which is \correct{predicted by galaxy merger simulations} \cite{Nakazato2024}.
Together, these findings \correct{reveal a diversity of past and current star formation activity among the different galaxies in the core of the protocluster in A2744, which can be connected with the merger-driven gas re-distribution, such as tidal gas stripping and resupplying, in shaping their current evolutionary stages.}

By combining the distribution of star-forming gas \correct{and its kinematics as revealed by \Ciium\ emission observed with ALMA (Figure \ref{fig:C2map}), together with star formation histories derived by JWST observations (Figure \ref{fig:SED})}, we find \correct{strong indications} that on-going galaxy-galaxy interactions play a dominant role on the evolution of the galaxies in the \quintet.
\correct{According to the recent change of their star formation activity ( Figure \ref{fig:SED}), the interaction among the individual components have occurred during the last $<50\,{\rm Myrs}$. For a typical \correctn{relative} velocity of $50$--$100\,{\rm km/s}$ (Figure \ref{fig:C2map}), this implies spatial scales of $<3$ -- $5\,{\rm kpc}$, which fit well with the relative projected distances between the components.
Although subject to projection effects, the spatial scales, the kinematics, and the star formation histories (SFHs) fit well within the proposed interaction scenario.}
This feature includes the picture that the merger-induced gas striping appears to have removed star-forming gas from galaxies such as YD7-E and ZD1 \correct{in the last $<50\,{\rm Myrs}$, resulting in the recent quenching of their star formation activities.}
In fact, \Ciium\ emission in the circumgalactic regions suggests the existence of massive \correct{($M_{\rm gas}=(2.3\pm0.2)\times10^{9}\,{\rm M_{\odot}}$)} neutral gas reservoir that could support the previous star-formation episodes of YD7-E and ZD1.
The stripped and the compressed gas are likely \correct{channeled} to the most massive galaxy within the system, YD4, where the supplied gas \correctn{appears to have fueled further refueling} of the star-formation, \correctn{as indicated} from its continuous star-formation activity and large gas reservoir.
At the same time, the compressed gas in the circumgalactic space likely triggered a burst of star-formation in \correct{a low mass}, very young galaxy, S1.

\correct{Cosmological simulations of gas-rich galaxy mergers at high redshifts shows that such processes can indeed occur in the early Universe, affecting the star formation of galaxies and producing complex star formation histories (see Methods).} These simulated merger events are finally found to lead to a rapid formation of a gas-rich\correct{, single} galaxy with stellar mass of $M_{\ast}\sim6\times10^{9}\,{\rm M_{\odot}}$ within \correctn{a time scale of} $\sim100\,{\rm Myr}$ \cite{Hashimoto2023}.
\correct{\correctn{Consistent with these theoretical predictions,} our analyses based on ALMA and JWST observations} reveal a dynamic cycle of gas stripping, redistribution, and recycling, leading to both quenching and ignition of star formation.
\correctn{Together, the simulation and observations present a unified picture in which major mergers within the core of a $z=7.9$ protocluster, with a total system mass of $\sim10^{10},{\rm M_{\odot}}$, are actively shaping the evolutionary pathways of its member galaxies during the epoch of reionization. Following the simulation, this system is expected to become $M_{\ast}>10^{10}\,{\rm M_{\odot}}$ in $\sim150\,{\rm Myr}$.}

This study provides the first direct observational constraints of how local 
environments in protoclusters can affect the evolution of massive galaxies in the early Universe.
\correct{These observations of the neutral gas reservoirs around one of the most extreme overdensities of galaxies in the early Universe reveal the radical impact these environments have on gas kinematics and, in turn, galaxy evolution at the earliest cosmic epochs.}

\newpage

\newpage

\newpage

\newpage


\begin{table}[h]
\centering
\begin{tabular}{lccccc}
\hline
ID & Stellar Mass & SFR$_{\rm 10Myr}$ &  \correct{Burstiness} & $M_{\rm H_2}$ & Status\\
& (log M$_\odot$) & (M$_\odot$ yr$^{-1}$) & (SFR$_{\rm 10Myr}$/SFR$_{\rm 100Myr}$) & (log M$_\odot$)  & \\

\hline
S1 & $7.6^{+0.4}_{-0.4}$ & $1.5^{+1.8}_{-0.7}$ & $3.5^{+10.7}_{-1.6}$ & $<8.8$ & starburst \\
YD1 & $8.4^{+0.3}_{-0.3}$ & $5.1^{+4.0}_{-2.5}$ & $2.0^{+4.5}_{-0.8}$ & $9.1\pm0.1$ & starburst \\
YD4 & $8.6^{+0.2}_{-0.3}$ & $5.0^{+5.4}_{-2.6}$ & $1.1^{+2.4}_{-0.4}$ & $9.1\pm0.1$& continuous \\
YD6 & $8.2^{+0.3}_{-0.2}$ & $1.0^{+1.3}_{-0.6}$ & $0.5^{+1.4}_{-0.3}$ & $\ast$ & continuous/quenching \\
YD7-E & $8.9^{+0.4}_{-0.1}$ & $1.1^{+5.8}_{-0.9}$ & $0.1^{+0.7}_{-0.1}$ & $<8.8$ & quenching \\
YD7-W & $8.3^{+0.3}_{-0.4}$ & $4.0^{+3.3}_{-1.9}$ & $2.1^{+5.2}_{-0.9}$ & $8.9\pm0.1$ & starburst \\
ZD1 & $8.2^{+0.2}_{-0.2}$ & $0.3^{+0.8}_{-0.2}$ & $0.3^{+0.9}_{-0.2}$ & $<8.8$ & quenching \\
CGM & -- & -- & -- & $9.3\pm0.3^{\ast\ast}$ & --\\
\hline
\end{tabular}
\caption{\correctn{Stellar mass ($M_*$), star formation rate averaged over the last 10 Myr (${\rm SFR_{10\,Myr}}$) and over the last 100 Myr (${\rm SFR_{100\,Myr}}$), burstiness defined as the ratio ${\rm SFR_{10\,Myr}}/{\rm SFR_{100\,Myr}}$, and molecular gas mass ($M_{\rm H_2}$) estimated from \Cii\ luminosities using the adopted conversion factor (see Methods).}\\
A gravitational magnification factor of $\mu=2$ is applied for \correct{correct} all quantities. \\
\correct{$\ast$ $M_{\rm gas}$ for YD4 includes YD6, as their \Ciium\ emission is unresolved in ALMA data.}
\correctn{$\ast\ast$  $M_{\rm gas}$ for CGM is estimated by \Cii\ luminosity seen in the YD1-YD4 bridge and the YD4-2 region.}
}
\label{tab:quintet_sfr}
\end{table}


\clearpage 

\bibliography{science_template} 
\bibliographystyle{sciencemag}


\section*{Acknowledgments}
This paper makes use of the following ALMA data: ADS/JAO.ALMA \#2023.1.00193.S and \#2023.1.01362.S. ALMA is a partnership of ESO (representing its member states), NSF (USA) and NINS (Japan), together with NRC (Canada), MOST and ASIAA (Taiwan), and KASI (Republic of Korea), in cooperation with the Republic of Chile. The Joint ALMA Observatory is operated by ESO, AUI/NRAO and NAOJ. In addition, publications from NA authors must include the standard NRAO acknowledgement: The National Radio Astronomy Observatory is a facility of the National Science Foundation operated under cooperative agreement by Associated Universities, Inc.

This work is based on observations made with the NASA/ESA/CSA James Webb Space Telescope. The data were obtained from the Mikulski Archive for Space Telescopes at the Space Telescope Science Institute, which is operated by the Association of Universities for Research in Astronomy, Inc., under NASA contract NAS 5-03127 for JWST. These observations are associated with program \#1324, \#2561, \#2756, and \# 4111.
Support for these programs was provided by NASA through a grant from the Space Telescope Science Institute, which is operated by the Association of Universities for Research in Astronomy, Inc., under NASA contract NAS 5-03127. 

Some of the data products presented herein were retrieved from the Dawn JWST Archive (DJA). DJA is an initiative of the Cosmic Dawn Center (DAWN), which is funded by the Danish National Research Foundation under grant DNRF140. 
\correct{Simulations were performed at Leibniz and Barcelona Supercomputing Centers.}

\paragraph*{Funding:}
Y.~F is supported by JSPS KAKENHI Grant Numbers JP22K21349 and JP23K13149.
Y.~N. acknowledges funding from JSPS KAKENHI grant No. 23KJ0728. 
M.~H. is supported by JSPS KAKENHI Grant No. 22H04939. 
A.K.I is supported by JSPS KAKENHI Grant No.21H04489, 23H00131, and 24H00002. 
T.H. is supported by JSPS KAKENHI Grant Numbers 23K22529, 25K00020, and 25H00661. 
CW acknowledges support from the Swiss State Secretariat for Education, Research and Innovation (SERI) under contract number MB22.00072.
\paragraph*{Author contributions:}
Y.~F. led the study, coordinated the ALMA observational programs, performed the data reduction and analysis, produced the figures, and wrote the manuscript.
Y.~N. performed and analyzed the cosmological simulations, and contributed to the interpretation of the results.
D.~C. developed the {\it FirstLight} simulation framework and provided a base-ground on the simulation analysis.
L.~C., S.~A., J.~A.~M., C.~B. P., L.~C., A.~C.~G., R.~M.~C.,and M.~P.~S. contributed to the JWST data analysis and interpretation.
T.~H., A.K.~I., K.~M., Y.~T., T.~B., H.~U., H.~M., M.~H., I.~M., W.~O., Y.~S., and Y.~W.~R. contributed to the ALMA data acquisition, reduction, and analysis.
N.~Y. provided theoretical context and interpretation for high-redshift galaxy evolution.
Y.~Z. and  C.~W. contributed to photometric and spectroscopic analysis of JWST data.
All authors discussed the results and contributed to improving the manuscript.
\paragraph*{Competing interests:} The authors declare that they have
no competing financial interests. Correspondence and requests for materials should be addressed to Y.F. (email: yoshinobu.fudamoto@gmail.com).

\paragraph*{Data and materials availability:}
The ALMA data we have used in this study can be accessible through the following data repository \url{https://github.com/yfudamoto/data-the_Quintet}.
The JWST data used in this study can be found in the DAWN JWST Archive \url{https://dawn-cph.github.io/dja/index.html}.


\newpage


\renewcommand{\thefigure}{S\arabic{figure}}
\renewcommand{\thetable}{S\arabic{table}}
\renewcommand{\theequation}{S\arabic{equation}}
\renewcommand{\thepage}{S\arabic{page}}
\setcounter{figure}{0}
\setcounter{table}{0}
\setcounter{equation}{0}
\setcounter{page}{1} 

\newpage


\section*{Methods}

\section{Cosmology}
Through out this analysis, we adopt a concordance cosmology with $\Omega_{\rm m}=0.3$, $\Omega_{\rm \Lambda}=0.7$, and $H_0=70\,{\rm km\,s^{-1}\,Mpc^{-1}}$.

\section{Targets}
The target of this study, A2744-{\em the Quintet} is located at ${\rm RA=3.603750\,deg}$, ${\rm Dec=-30.382250\,deg}$ behind the gravitationally lensed galaxy cluster Abell-2744 \cite{Abell1989}.
YD1, YD4, YD6 and YD7 in A2744-{\em the Quintet} were observed and detected as a part of Hubble Frontier Fields Survey \cite{Lotz2017} and reported in several papers accompanied with Spitzer/IRAC observations \cite{Laporte2014,Zheng2014,Ishigaki2016}.
Later, JWST NIRSpec spectroscopy identified that  A2744-{\em the Quintet} is a part of the galaxy protocluster at $z\sim7.9$, A2744-zp9OD \cite{Morishita2023}.
Several JWST NIRSpec observations were performed including NIRSpec integral field spectroscopy (IFS) \cite{Hashimoto2023, Venturi2024} and micro shutter array (MSA) observations \cite{Witten2025}.

These observations find that the \quintet\ consists of at least $5$  galaxies, namely YD1, S1, YD4, YD7,
and ZD1 (see also \cite{Venturi2024} for more detailed components).
Additionally, clear sub-components of YD7 is separately named as YD7-E and YD7-W in \cite{Hashimoto2023}.
Similarly, extended sub-component of YD4 is named as YD6. 
\correct{YD1 is separately named as YD1-E and YD1-W based on their rest-frame UV morphology and \Oiiilong\ emission line detections \cite{Venturi2024}.}
For each components, we adopt spectroscopic redshift determined in \cite{Hashimoto2023} for YD1, YD4, YD7, and S1.
The redshift of ZD1 is adopted from \cite{Venturi2024}.
We apply redshifts of their host components to their sub-components.
Names of the components are summarized in Figure \ref{fig:quintet}.

\section{ALMA observations and far-infrared analysis}
\subsection{Observations and data reduction}

\subsubsection*{[C{\sc II}]$\,158\rm \mu m$ emission line}
To observe the \Ciium\ emission line of the target galaxies, {\em A2744-the Quintet}, the ALMA band-6 observations were performed during ALMA cycle-10. 
The observations have targeted the \Ciium\ emission line that is redshifted to the observed frequency of $\nu_{\rm obs}\sim214.2\,{\rm GHz}$.
For studying the \Ciium\ emission in this paper, we combined two ALMA observations performed separately; program IDs of \#2023.1.00193.S and \#2023.1.01362.S.

\#2023.1.00193.S  (PI: Y. Fudamoto) was performed between 2nd and 12th of June, 2024.
Weather conditions were good with precipitable water vapors (PWVs) between $1.3$ to $2.0\,{\rm mm}$. The correlator was configured to cover $211.43$ -- $215.18\,{\rm GHz}$ and $226.31$ -- $230.06\,{\rm GHz}$ using four spectral windows (SPWs). Each SPW was set to the Frequency Division Mode (FDM) with a $1.875\,{\rm GHz}$ bandwidth and a $7.8125\,{\rm MHz}$ ($10.93\,{\rm km/s}$ at $214.2\,{\rm GHz}$) resolution. 
The median synthesized beam is $0\farcs41\times0\farcs36$ in a full width at half maximum (\correct{FWHM}) with the position angle (PA) of $30.4\,{\rm degree}$ when the \texttt{NATURAL} weighting is applied.
The total on-source time was $11.4\,{\rm hours}$.

\#2023.1.01362.S  (PI: Y. Tamura) was performed between 8th of January and 28th of April, 2024.
Weather conditions were good with PWVs between $1.7$ to $3.3\,{\rm mm}$.
The correlator was set to the FDM, covering $213.53$ -- $216.39\,{\rm GHz}$ and $226.06$ -- $229.79\,{\rm GHz}$. One of the SPWs that covers the redshifted \Ciium\ emission frequency was set to the 4-bit mode with a bandwidth of $937.5\,{\rm GHz}$ and the rest of each spectral window was set to the 2-bit mode with a bandwidth of $1.875\,{\rm GHz}$ bandwidth and a $3.906\,{\rm MHz}$ ($5.44\,{\rm km/s}$ at $214.2\,{\rm GHz}$) resolution. 
The median synthesized beam is $1\farcs3\times1\farcs1$ in a \correct{FWHM} with the PA of $86.0\,{\rm degree}$ when the \texttt{NATURAL} weighting is applied.
\correct{The observation employed seven pointing to map a wide are of $80^{\prime\prime}\times40^{\prime\prime}$ covering a galaxy over-density A2744-7p9OD \cite{Morishita2023}.}
Thus, we only used three pointing out of the entire seven pointing that include the target galaxies \correct{within their half primary beam widths}.

All the data were pipeline-calibrated during the quality assurance procedure by the ALMA observatory. We reproduced the calibrated measurement sets (MSs) using the observatory provided script ``\texttt{ScroptForPI.py}'' using the CASA software version 6.5.4.9 and the pipeline version 2023.1.0.124.
We find that the data and the calibration quality were excellent and the atmospheric transmission is good over the entire frequency ranges, resulting in the stable noise levels for all channels.
Therefore, we did not perform any manual flagging in addition to the pipeline calibration. 

From the calibrated data of \#2023.1.01362.S, we split three pointing that cover sky regions surrounding the {\em A2744-the Quintet} using \texttt{split} task of CASA.
The calibrated MSs of the three pointing of \#2023.1.01362.S and the calibrated MS of \#2023.1.00193.S are then concatenated using \texttt{concat} task of CASA. 

To create emission line data cube, we used \texttt{tclean} task of CASA. We applied channel width of $15\,{\rm km/s}$ and set the rest sky frequency of $214.18\,{\rm GHz}$ determined as the redshifted \Ciium\ frequency from the target galaxy at $z=7.87$ \cite{Hashimoto2023}.
We performed a synthesized beam deconvolution using a stopping threshold of $2\times$ the typical root-mean-square (RMS) of background measured from a channel of the ``dirty'' image before the deconvolution. We also applied a maximum iteration count of $10^6\,{\rm times}$, which was not reached during the beam deconvolution process. 
\correct{To properly deconvolve the synthesized beam from our target that include both extended and compact components, we applied \texttt{multiscale} deconvolution algorism}. 
The \texttt{Natural} weighting scheme, and \texttt{standard} gridder were applied.
We do not perform a continuum subtraction from the data cube to avoid any additional uncertainty. The dust continuum emission of the target galaxies are faint as described below and do not affect emission line measurements.

\subsubsection*{Dust continuum}

To examine the dust continuum in ALMA Band-6 data, we use an ancillary ALMA archival data, \#2018.1.01332.S in addition to the observations performed for the \Ciium\ observations above.
\correct{\#2018.1.01332.S} (PI: N. Laporte) was performed between 1st and 2nd of May, 2019.
Weather conditions were excellent with PWVs between $0.7$ to $1.3\,{\rm mm}$. The correlator was configured to cover  $259.20$ -- $261.12\,{\rm GHz}$ and $245.39$ -- $247.2\,{\rm GHz}$ using four SPWs. One SPW centering to $262.06\,{\rm GHz}$ was set to the FDM with a $1.875\,{\rm GHz}$ bandwidth and a $7.8125\,{\rm MHz}$ ($8.93\,{\rm km/s}$ at $262.06\,{\rm GHz}$) resolution and other three SPWs were set to the Time Division Mode (TDM) with a $2\,{\rm GHz}$ bandwidth.
The median synthesized beam is $0\farcs 55\times0\farcs 47$ in a FHWM with the PA of $-85.2\,{\rm degree}$ when the \texttt{NATURAL} weighting is applied.
The total on-source time was $5.2\,{\rm hours}$. The data were pipeline-calibrated during the quality assurance procedure of the ALMA observatory, and we reproduced the calibrated MS using the observatory provided script ``\texttt{ScroptForPI.py}'' using the CASA software version 5.4.0-70 and the pipeline version 42254.

We first created MSs that have frequency coverages only over the dust continuum. We conservatively created the continuum MSs by removing channels that corresponds to $\pm3\sigma$ widths of the detected \Ciium\ emission lines.
This procedure corresponds to removing channels between $213.8$ -- $214.5\,\rm{GHz}$ both from \#2023.1.00193.S and from \#2023.1.01362.S.
We visually inspected that the removing frequency range can securely remove \Ciium\ emission line.
We then concatenated all MSs using CASA task \texttt{concat} and applied $30\,\rm{s}$ time averaging to reduce the data volume.
By concatenating three data sets, dust continuum MS covers the observed-frame frequency of $\nu_{\rm obs}=213.53$ -- $261.12\,{\rm GHz}$ (or $\lambda_{\rm rest} = 129$ -- $158\,\rm{\mu m}$ in the rest-frame of {\em the Quintet} at $z=7.9$). Therefore, the concatenated data represents dust continuum emission at the average frequency of $\nu_{\rm obs}=237.2\,{\rm GHz}$ or the average rest-frame wavelength of $\lambda_{\rm rest}=142\,{\rm \mu m}$.

The dust continuum map was made using multi-frequency synthesis (\texttt{mfs}) mode of the \texttt{tclean}. We performed a deconvolution targeting the $2\times$RMS of the background measured in the dirty image.
We also applied the maximum iteration count of $5000\,{\rm times}$, which was not reached during the beam deconvolution process. The \texttt{Hogbom} deconvolution algorism, the \texttt{Natural} weighting scheme, and the \texttt{standard} gridder were applied.
Finally, we made two continuum images; one employing tapering and one without tapering.
The continuum image without tapering have a synthesized beam FWHM of $0\farcs56\times0\farcs50$ with a PA of $87.2\,\rm{deg}$ and reached a background RMS of $3.6\,{\rm \mu Jy/beam}$. In the another continuum image, we applied a $uv$-taper using \texttt{outertaper}$=0\farcs4$. As a result, the tapered continuum image has a synthesized beam FWHM of $0\farcs88\times0\farcs77$ with a PA of $87.5\,\rm{deg}$ and reached a background RMS of $4.0\,{\rm \mu Jy/beam}$.

\subsection{Results}
\subsubsection*{Detected \Ciium\ emission}
\begin{figure}
    \centering
    \includegraphics[width=1.0\linewidth]{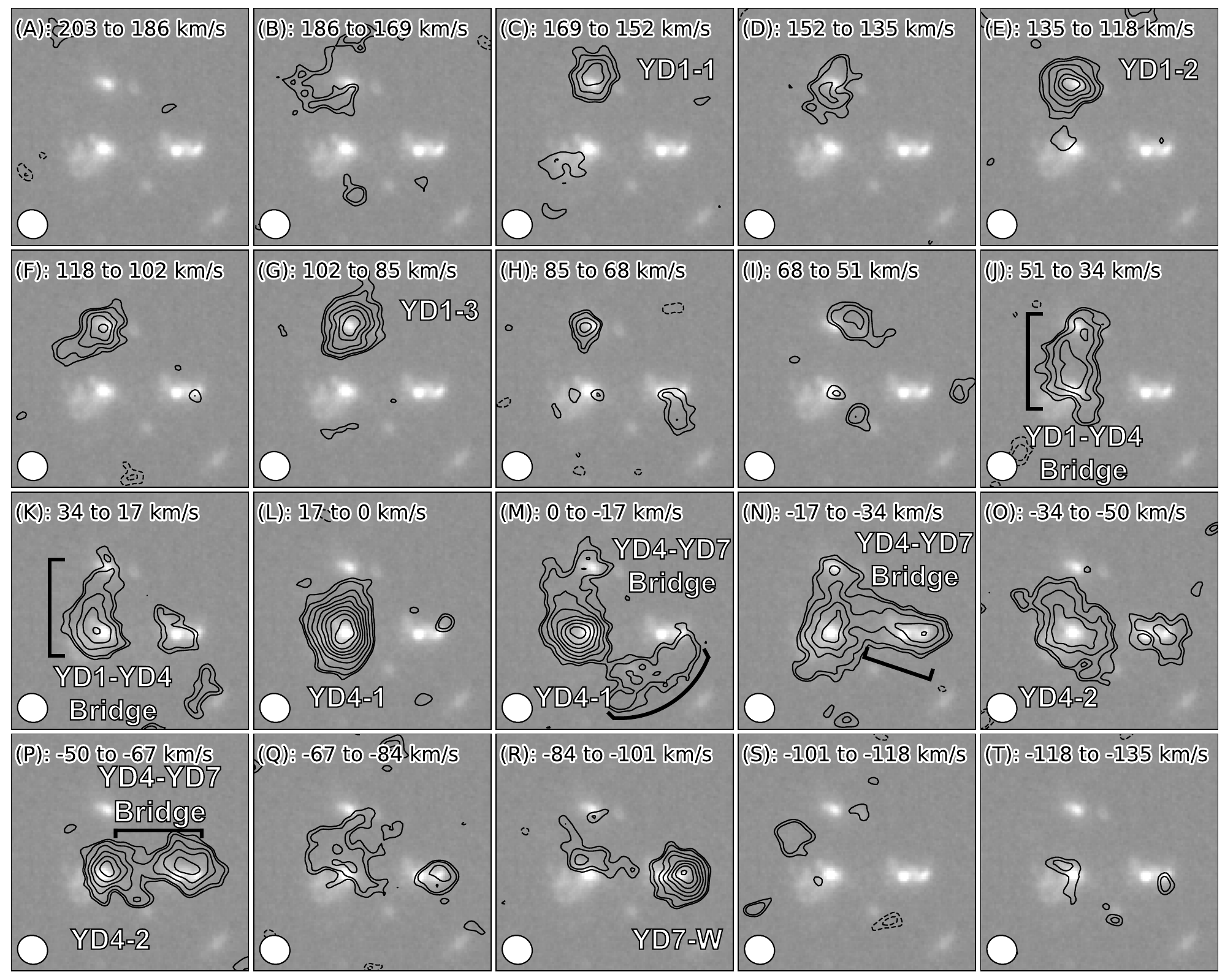}
    \caption{{\bf Velocity channel map of \Ciium\ emission lines of \quintet.}
    \correct{The background image shows the stacked image of JWST's long wavelength filters.}
    The velocity range is labeled at the top of each panel. Clumps and bridges are labeled in the maps. The synthesized beam is shown at the lower left corner of each map. \correct{Contours start from $2\,\sigma$ and end at $9\,\sigma$ with a step of $1\,\sigma$.}
    }
    \label{fig:velmap}
\end{figure}

\begin{figure}
    \centering
    \includegraphics[width=1.0\linewidth]{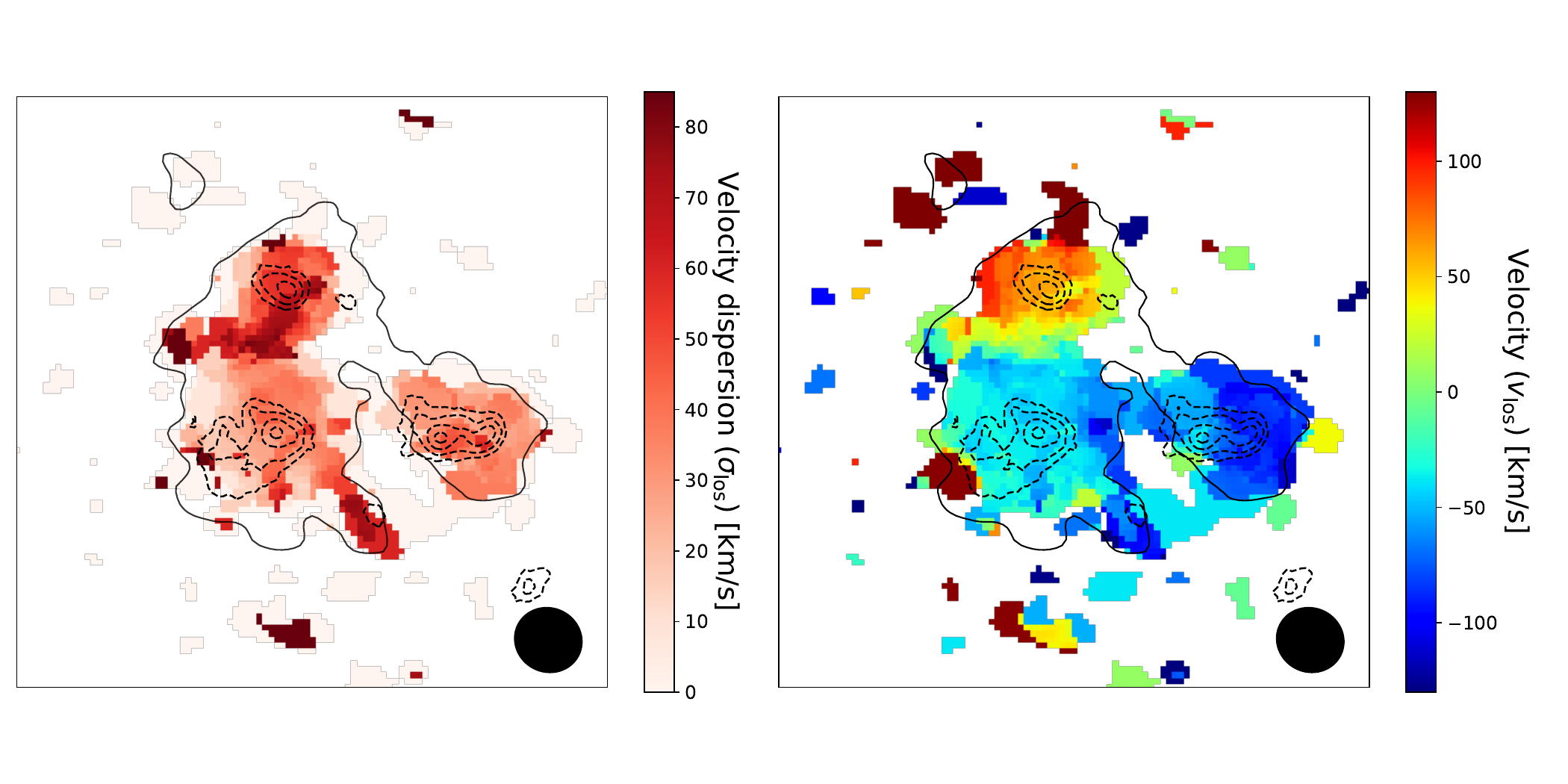}
    \caption{\correct{{\bf The projected velocity and velocity dispersion field of the \Ciium.}
    Left panel shows the line of sight velocity dispersion of \Cii\ emission, while the line of sight velocity of \Cii\ emission is shown in right panel.
    The black solid contour shows the 2 sigma level of \Ciium\ emission.
    Dashed contours show $6, 12, 24, 48\,\sigma$ of JWST LW stacked image. Black ellipses at the lower right corners show the synthesized beam size of ALMA observations.}}
    \label{fig:velocity}
\end{figure}

The \Ciium\ emission line is significantly detected in multiple components across the \quintet\ over multiple velocity channels (Figure \ref{fig:velmap}). In particular, three galaxy components (YD1, YD4, YD7-W) show robust detections at SNRs of $6$ -- $9$ in some of the velocity channel maps. Furthermore, the spatial distribution of the \Ciium\ emission extends well beyond the rest-frame UV and optical components of these galaxies, forming filamentary structures that connect multiple galaxies in several directions, which we call ``bridge'' hereafter.
The bridge features suggest the existence of dynamical interactions between the galaxies and gas flows within the system \cite{Izumi2024,Zanella2024,Sugahara2025}.
\correct{The velocity dispersion of the \Cii\ emission in the bridge structures show $\sim2\times$ higher than the bright clumps, further supporting the scenario that the bridge structures are shaped by  dynamical interactions (Figure \ref{fig:velocity}).}
In contrast, other member galaxy components (ZD1, YD7-E, S1) show little or no detectable \Ciium\ emission, despite some of these components are brightly detected in the rest-frame UV and optical continuum and/or emission lines \cite{Hashimoto2023, Venturi2024}.
We describe the \Ciium\ emission line properties of each galaxy below, ordered by \correctn{decreasing \Ciium-based redshift}.
In the following description, we use the \Ciium\ redshift of $z_{\rm [CII]}=7.8744$ as a reference value, which is the redshift of YD4, \correct{the \Cii\ brightest} galaxy component in the \quintet.

{\bf YD1:} The \Ciium\ emission in YD1 has two to three distinct clumps along the line of sight.
The \Cii\ line spans a velocity range of $\Delta v_{\rm [CII]}\sim120\,{\rm km/s}$ from $\sim+169$ to $\sim+51\,{\rm km/s}$ relative to YD4.
The total \Cii\ luminosity is estimated by summing the fluxes of these clumps  (corresponding to the panels C -- H in Figure \ref{fig:velmap}), yielding  $(8.2\pm0.6)\times10^7\,{\rm L_{\odot}}$.
The \Cii\ redshift is measured to be $z_{\rm [CII]}=7.8778$ from the peak frequency of the detected \Cii\ emission.
The \Cii\ redshift is consistent with the redshift derived from the \Oiiiboth\ emission lines \cite{Hashimoto2023}, confirming that both emission lines trace the same galaxy component.

{\bf YD1-YD4 Bridge:} In the velocity range between $+51$ to $+17\,{\rm km/s}$ relative to YD4, we identify a spatially extended \Ciium\ structure located betwen YD1 and YD4, spanning $\sim5.8\,{\rm kpc}$ in projection.
Based on the spatial distribution and the velocity structure, we refere to this feature as the YD1-YD4 bridge.
There is no underlying stellar comonent detected in the rest-frame UV and optical images,
indicating that this bridge components smoothly traces a tidal gas stream formed by the interaction between YD1 and YD4.
\correct{Given the stellar mass ratio of 1:0.45 between YD4+YD6 and YD1 as well as the larger gas mass in YD4}, the gas in the bridge is presumably accreting toward the deeper gravitational potential of YD4.
The \Ciium\ luminosity of this structure is measured to be $(5.6\pm0.7)\times10^{7}\,{\rm L_{\odot}}$ by summing the \Cii\ fluxes detected in panels I -- K of figure \ref{fig:velmap}.

{\bf YD4:}
The \Ciium\ emission associated with YD4 has two peaks along the line of sight.
The brightest \Cii\ emitting component (YD4-1) is detected at $\nu_{\rm obs}=214.156\,{\rm GHz}$ (Panel L of Figure \ref{fig:velmap}).
Its \Cii\ redshift of $z_{\rm [CII]}=7.8744$ \correctn{differs by only $\Delta v=+60\,{\rm km/s}$} from the redshift measured from \Oiiiboth\ lines in the NIRSpec high-resolution spectrum, \correctn{which has a spectral resolution of $\sim100\,{\rm km/s}$ \cite{Hashimoto2023}}.
\correctn{The consistent redshifts between \Cii\ emission and optical emission indicate} that YD4-1 seen in \Cii\ emission corresponds to the same star-forming component of YD4 traced by those rest-frame optical emission lines.
The \Ciium\ luminosity of YD4-1 is estimated to be $(9.3\pm 0.6)\times10^7\,{\rm L_{\odot}}$ by summing the \Cii\ fluxes over the velocity range of $17$ to $-34\,{\rm km/s}$ (Panel L to N in Figure \ref{fig:velmap}).
The beam deconvolved size of YD4-1 is estaimted to be $1\farcs0\times0\farcs5$ (or $\sim4.9\times2.5\,{\rm kpc}$), indicating that YD4-1 is significantly more extended than the stellar component of YD4.
A secondary fainter, blueshifted component (YD4-2) is also significantly detected over the velocity range of $-50$ to $-67\,{\rm km/s}$ (Panel P of Figure \ref{fig:velmap}).
YD4-2 is extended along a line of sight over the velocity range of $-34$ to $-84\,{\rm km/s}$ (Panel O and Q in Figure \ref{fig:velmap}) with a peak frequency of $\nu_{\rm obs}= 214.204\,{\rm GHz}$ or ($z_{\rm [CII]}=7.8724$).
The \Ciium\ luminosity of YD4-2 is estimated to be $(8.0\pm 0.7)\times10^7\,{\rm L_{\odot}}$ by summing the \Cii\ fluxes over the velocity range.
\correct{No clear counter part of YD4-2 is detected in previous JWST observations of rest-frame optical emission lines and have the same line-of-sight velocity as the YD4-YD7 bridge (see following). These features suggest that YD4-2 is a diffuse gas components connected with the YD4-YD7 bridge, and lacking clear sign of significant on-going star formation.}

{\bf YD4-YD7 Bridge:} Two to three spatially extended \Ciium\ emission regions are detected between YD4 and YD7, forming filamentary bridges that connect YD4 and YD7 (Panels M, N, and P in Figure \ref{fig:velmap}).
These bridges are all blueshifted with respect to YD4-1 between $-17$ to $-67\,{\rm km/s}$.
One of the bridge connects YD4-2 and YD7 at the most blue-shifted end of the detected velocity range (Panel P of Figure \ref{fig:velmap}), suggesting a continuous tidal gas stream between two galaxies.
As with the YD1-YD4 bridge, no underlying stellar component is detected in the rest-frame UV and optical images.
The spatial structure of the bridge, particularly its connection between YD7 and YD4-2, indicates that it is a tidal gas structure formed during the merger, redistributing gas between YD4 and YD7.

{\bf YD7-W:} Several \Ciium\ clumps are detected toward YD7 along the line of sight with the velocity range between $-34$ to $-100\,{\rm km/s}$ (panel N to R in Figure \ref{fig:velmap}).
The brightest \Cii\ clumps is spatially coincident with YD7-W and lies within the velocity range between $-84$ to $-101\,{\rm km/s}$ (panel R of Figure \ref{fig:velmap}), corresponding to the \Cii-based redshift of $z_{\rm [CII]}=7.8714$. 
This  \Cii\ redshift is slightly ($\sim25\,{\rm km/s}$) blueshifted relative to the \Oiiiboth-based redshift of YD7-W ($z_{\rm [OIII]}=7.8721\pm0.0001$) reported in \cite{Hashimoto2023}.
The \Ciium\ luminosity of YD7-W is estimated to be $(2.3\pm 0.3)\times10^7\,{\rm L_{\odot}}$ by summing fluxes over the velocity range of $-67$ to $-101\,{\rm km/s}$ (Panel Q and R in Figure \ref{fig:velmap}).
In the velocity range of $-17$ -- $-67\,{\rm km/s}$, additional significant \Ciium\ emission components are detected. However, their spatial peaks are not clearly associated with YD7-W.
While the exact nature of these feature is uncertain, we tentatively interpret them as a part of the YD4-YD7 bridge. Alternatively, they may be partially associated with YD7-E and/or unresolved substructures of YD7 along the line of sight.
The total \Ciium\ luminosity of these intermediate-velocity components is estimated to be $(5.6\pm 0.6)\times10^7\,{\rm L_{\odot}}$ based on fluxes in Panel N to P of Figure \ref{fig:velmap}.

\correct{Finally, we estimate molecular gas masses ($M_{\rm H_2}$) using the measured \Ciium\ luminosities, assuming that the \Cii\ emission traces predominantly molecular gas regions.
We note that this conversion is subject to uncertainties due to its dependence on local ISM conditions and metallicity, while these estimates offer a view of the cold gas content in each galaxy. To do this, we assume a conversion factor measured in \cite{Zanella2018} and report the results in Table \ref{tab:quintet_sfr}.}

\subsubsection*{\Ciium\ emission non-detected components}

\correct{ZD1 and S1 do not show clear \Ciium\ emission peaks.
Although some of the neighboring \Cii\ signals may overlap with these components, we conclude they are non-detected in \Cii\ due to the lack of  significant peaks.
From the RMS of moment-0 map, we calculate $3\,\sigma$ upper limits of \Ciium\ luminosity of $L_{\rm [CII]}<2.2\times10^7\,{\rm L_{\odot}}$ for ZD1 and S1.}
YD7-E, a UV and optically luminous subcomponent of YD7, also lacks a distinct \Ciium\ peak.
Although no clear emission is directly associated with YD7-E, some of the \Cii\ associated with YD7-E lies along the projected extent of the YD4-YD7 bridge.
It is therefore possible that gas associated with YD7-E is blended into the extended tidal structure within YD4-YD7 bridge, making it difficult to isolate. However, the faintness of the \Ciium\ is clearly shown by the non-detection of clear, distinct \Cii\ peak colocated with YD7-E, which is the most UV/optcally luminous galaxy component in the \quintet..

\subsubsection*{Dust continuum}
\begin{figure} 
	\centering
	\includegraphics[width=0.8\textwidth]{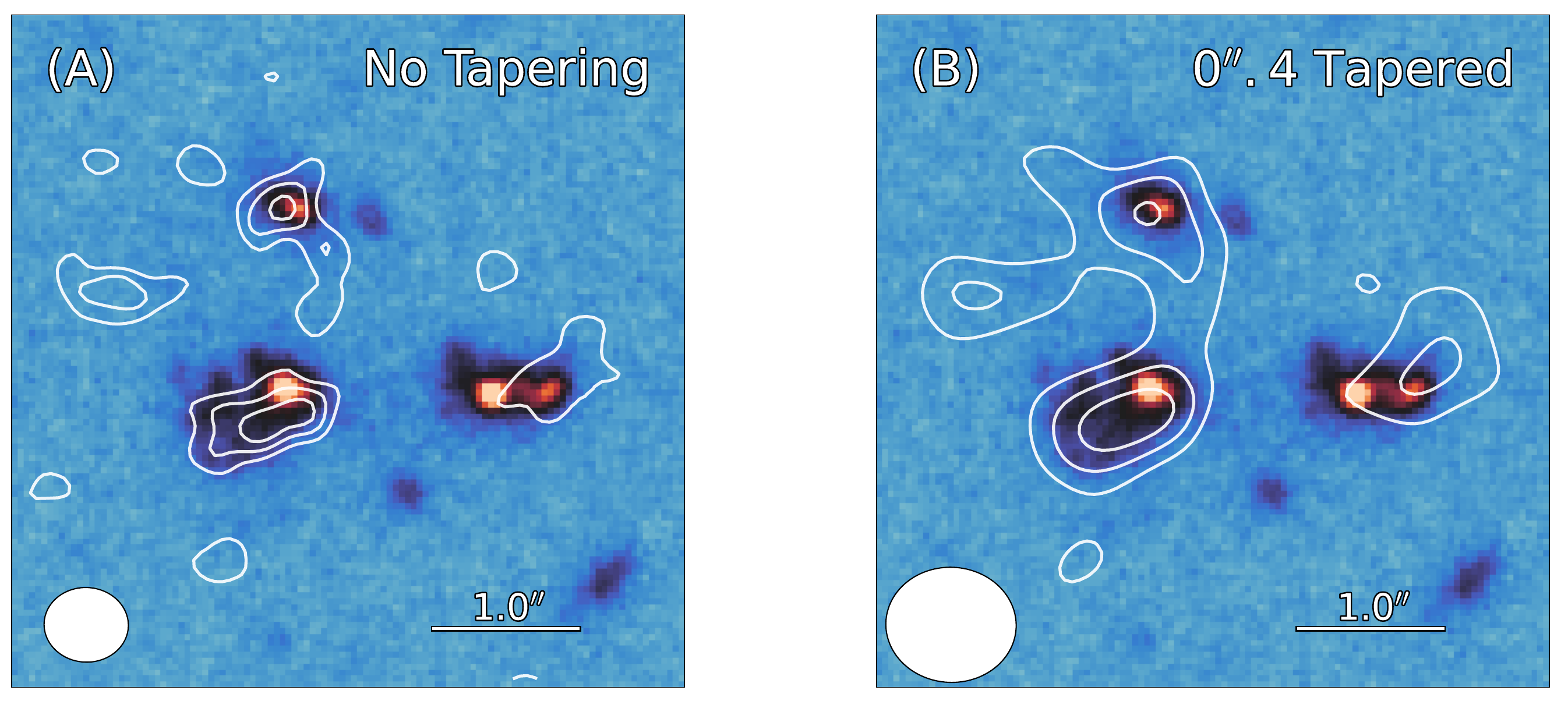} 

	\caption{\textbf{The rest-frame $\mathbf{142\,{\rm \mu m}}$ dust           continuum emission of A2744-the Quintet at $z=7.88$:}
	   {\em Panel (A)}: The backgound image shows a stacked JWST image at wavelength longer than $2.7\,{\rm \mu m}$. White contours show dust continuum emission obtained by ALMA. Contours start from $2\,\sigma$ with a step of $1\,\sigma$. The ellipse at the lower left corner shows the synthesized beam FWHM of the dust continuum map.
       Dust continua are detected from three member galaxies of {\em the Quintet} (YD4, YD1, YD7-W). 
       {\em Panel (B)}: Same as the panel (A) but the dust continuum image is produced by applying a $0\farcs4$ tapering.
        }
	\label{fig:continuum} 
\end{figure}

\correct{The dust continuum emission is detected in three members of} \quintet; YD1, YD4, and YD7-W (Figure \ref{fig:continuum}).
In particular,  from YD4, we find a dust continuum detected with a peak SNR of $4.6\,\sigma$ in the non-tapered image and $4.5\,\sigma$ in the $0\farcs4$-tapered image. From YD1, we find $4.2\,\sigma$ detections both from the non-tapered and from the $0\farcs4$-tapered image. 
YD7-W has a tentative $2.8\,\sigma$ signal in the non-tapered image. The tentative signal from YD7-W become $3.3\,\sigma$ in the $0\farcs4$-tapered image, suggesting that the dust continuum in YD7-W is extended.
To test whether different tapering parameters will make the continuum signal from YD7-W more significant, we applied tapering starting from $0\farcs3$ fo $0\farcs8$ with a step of $0\farcs05$. We find $0\farcs4$ tapering gives the highest SNR.
\correct{Within $1\,\sigma$ uncertainty, the peak signals of dust continua are consistent with the ones reported in the previous shallower data  \cite{Hashimoto2023}.}

We also find continuum signal extended from the host galaxies toward 
circumgalactic space between member galaxies of A2744-{\em the Quintet}. However, these signals are only tentative, while similarly extended \Ciium\ emission qualitatively consistent with these extended signals.

Flux densities of dust continua are measured from peak pixel values using spatially unresolved data produced by applying $0\farcs6$ tapering (Table \ref{tab:dust}). The measured dust continuum fluxes are found to be well consistent with the aperture flux measurement using an aperture with a radius of $1\farcs0$.
In contrast, we find that flux measurements using 2-D Gaussian fitting (e.g., CASA task \texttt{imfit}) produces $\sim2\,\times$ larger values. The discrepancy would be due to the over-estimations of source sizes in 2-D Gaussian fitting as multiple dust continuum detected sources are located close together. 
The over-estimations of sizes are also shown by the beam deconvolved source sizes of $\sim1^{\prime\prime}$ found by \texttt{imfit}, which is beyond the JWST's rest-frame optical detections. Thus, we do not use results from 2-D Gaussian fitting in this study.
\correct{We note that the previous study using shallower data  reported $3$ -- $4\times$ larger dust continuum fluxes using 2-D Gaussian fit \cite{Hashimoto2023}. The higher fluxes in the previous study would also come from the over-estimation of source size due to the complex morphology as tested here.}
For non-detected galaxies, we conservatively estimate the $3\,\sigma$ upper limits by calculating $3\times$ the background RMS of $0\farcs6$ tapered dust continuum image.

Using the measured dust continuum fluxes, we estimate IR luminosity ($L_{\odot}$) of member galaxies of \quintet. As an FIR SED, we assume a modified black body with a single  dust temperature of $T_{\rm d}=50{\rm K}$ and a dust emissivity index of $\beta_{\rm d}=1.5$. The FIR SEDs are scaled to the measured fluxes and are integrated over the rest-frame wavelength range between $\lambda_{\rm rest}=8$ -- $1000\,{\rm \mu m}$.
We estimate dust masses using the same FIR SED and assuming dust mass absorption coefficient of $\kappa = \kappa_0\,(\nu/\nu_0)^{\beta_{\rm d}}$, where $\kappa_0=10\,{\rm cm^2\,g^{-1}}$ at $\lambda_{\rm rest} = 250\,{\rm \mu m}$ \cite{Hildebrand1983}.
These estimated properties are summarized in Table \ref{tab:dust}.

\section{JWST Observations and rest-frame UV/optical analysis}

\subsection{JWST observations and photometry}

JWST's \correct{Near Infrared Camera (NIRCam)} imaging of {\em A2744-the Quintet} was obtained as part of the extensive campaign targeting the A2744 galaxy cluster, performed across multiple JWST programs; GO-1324 (PI: T. Treu), GO-2561 (PI: I. Labbe, R. Bezanson), GO-2756 (PI: C. Wenlei, H. Williams) and GO-4111 (PI: K. Suess).
These programs provided imaging in all available broad- and medium-band filters during JWST Cycle-1 to Cycle-3. 
Detailed descriptions of the images and data processing are described in \cite{Roberts-Borsani2022,Weaver2024, Suess2024}.
We used a total of 19-band JWST images retrieved from the DAWN JWST Archive \cite{Valentino2023} to perform  photometry of the \quintet.

Prior to photometric measurements, all images were point spread function (PSF)-matched to the F480M filter \correct{(i.e., the filter with the lowest resolution PSF with FWHM $\sim0\farcs16$)}. The PSF of each image was created using the \texttt{stpsf} tool \cite{Perrin2014}, and convolution kernels are generated using the \texttt{pypher} package \cite{boucaud2016}. Based on the PSF-matched images, we performed both circular aperture photometry on each bright clumps and pixel-by-pixel photometry. 

For the clump-based aperture photometry, we adopted circular apertures with a fixed aperture radius of $0\farcs1$. \correctn{To correct for the PSF assuming point sources, all fluxes were multiplied by a factor of 2.2. This correction factor was derived using a $0\farcs1$ radius circular aperture and the F480M PSF generated with \texttt{stpsf}.}
Photometric uncertainties were estimated by randomly placing 5000 apertures of the same size ($r=0\farcs1$) across each image. 

\subsection{Lack of active galactic nucleus}

We investigate the \correct{possible} presence of AGNs in member galaxies of the \quintet\ from their morphology in the rest-frame UV continuum and their emission lines. 
In the rest-frame UV image, most of the member galaxies have spatially extended morphology, suggesting that there is no dominant rest-UV emission that originates from point-like sources such as AGNs \correct{(Figure \ref{fig:quintet}, bottom left panel)}. In the rest-UV images, YD4 and YD7-E have bright and compact emission.
However, YD4 does not have any signs of broad Balmer lines, especially in its H$\beta$ line reported in the previous study \cite{Venturi2024}. \correct{Also YD4 do not show any significant emission lines in rest-frame UV \cite{Witten2025}.} YD7-E does not have significant rest-frame optical emission lines \correct{as seen in its multi-wavelength photometry and its spectra \cite{Hashimoto2023}}, showing no narrow or broad line region.
These interpretations of lack of AGN activities are consistent with other studies based on spectroscopy \cite{Venturi2024}. Therefore, we conclude that member galaxies of the \quintet\ do not host AGNs and their emission is dominated by star-formation activities.

\subsection{Spectral energy distribution (SED) modeling}

To infer galaxy properties and star-formation histories (SFHs) of member galaxies of the \quintet, we performed SED fitting using the circular aperture photometry.
We used the SED fitting code \texttt{PROSPECTOR} \cite{Johnson2021}. 
F115W photometry was excluded from the fits due to the Ly$\alpha$ damping wing absorption \cite{Chen2024,Witten2025}, which currently lacks reliable modeling in SED fitting frameworks.
\texttt{PROSPECTOR} models nebular line and continuum emission using FSPS \cite{Byler2017} combined with \texttt{CLOUDY} \cite{Ferland2013} photoionization models. Dust attenuation is treated using the two-component model of \cite{Charlot2000}, with a dust index allowed to vary between $-1.2$ and $0.4$. We fixed the dust attenuation for stars younger than 10 Myr (dust1) through a functional dependence on dust2, and allowed the dust1 fraction to vary freely with a truncated normal prior centered at 1.0.
We adopted a non-parametric SFH using seven age bins spanning from 0 to $\sim$0.46 Myr: [0–10$^7$], [10$^7$–$2\times10^7$], [$2\times10^7$–$5\times10^7$], [$5\times10^7$–$10^8$], [$10^8$–$1.5\times10^8$], [$1.5\times10^8$–$3\times10^8$], and [$3\times10^8$–$4.6\times10^8$] years. The relative SFRs between adjacent bins are regularized by a continuity prior following a Student's $t$-distribution with scale $\sigma=0.3$ and degree of freedom $\nu=2$. We used a fixed Chabrier IMF \cite{Chabrier2003} and configured other parameters following standard prescriptions in \texttt{PROSPECTOR}. 

\subsubsection*{SED fitting Results}
Results of the SED fitting are summarized in Table \ref{tab:quintet_sfr}.
While our estimation might miss some of the components due to using small size apertures, we find that the magnification corrected stellar masses of the \quintet\ member span a range of $M_{\star}\sim10^{7.6}$ to $10^{8.9}\,{\rm M_{\odot}}$, with star formation rates (SFRs) ranging from ${\rm SFR_{10}}=0.3$ to $5.1\,{\rm M_{\odot}\,yr^{-1}}$ where ${\rm SFR_{10}}$ is the magnification corrected SFR average over the last $10\,{\rm Myr}$ of the SFH.
These results are consistent with the previous studies \cite{Hashimoto2023,Witten2025}.
Furthermore, by performing SED fittings to multiple components of member galaxies, we find a broad diversity in star formation activity and star formation histories among the \quintet\ member galaxies.
These contrasting properties reveal the diversity of evolutionary stages in the system and show their formation histories inprinted in the SED fitting results. We detail the SED fitting results below.

\correct{We classify star formation modes based on the median value of the ratio between the recent SFR (averaged over the last $10\,{\rm Myr}$) and the average SFR over the past $100\,{\rm Myr}$.
Galaxies with a median ${\rm SFR_{10\,Myr}/SFR_{100\,Myr}} < 1$ are classified as quenching, those with ratios around unity as continuous, and those with ratios greater than 2 as starburst galaxies (Table \ref{tab:quintet_sfr}).}

\begin{figure}
    \centering
    \includegraphics[width=1.0\linewidth]{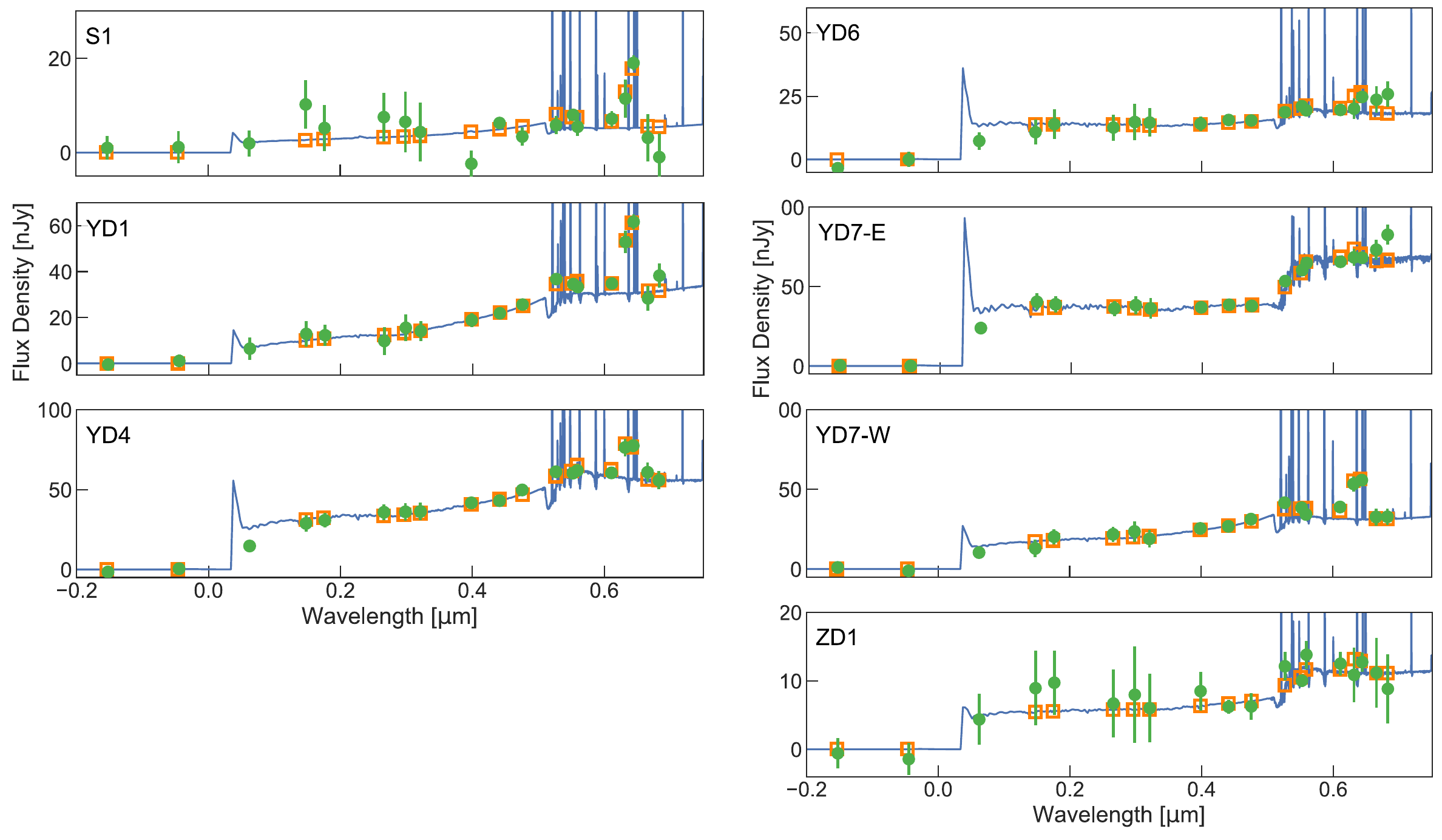}
    \caption{{\bf Obtained photometry and the SED fitting results of the \quintet\ member galaxies} Green points with errorbars show the observed photometry and the blue lines and orange squares show the fitted model SEDs and photometry, respectively.}
    \label{fig:SEDspectra}
\end{figure}

YD1, S1, and YD7-W have a burst-like star-formation histories, characterized by rapidly rising SFRs in the last $10\,{\rm Myr}$ of $1.5$ to $5.1\,{\rm M_{\odot}\,yr^{-1}}$. The specific star-formation rates (sSFRs) of these galaxies ranges from $20$ to $38\,{\rm Gyr^{-1}}$, indicating that these galaxies are experiencing a starburst phase. Among these galaxies, S1 has the most extreme bursty behavior, as is evident from its SED, which shows weak stellar continuum but is dominated by strong emission line (Figure \ref{fig:SEDspectra}).

\correct{YD4 shows an extended and rising star-formation history. We note that the derived star-formation history shows different features compared with previous studies \cite{Witten2025}, which suggest YD4 has a “rejuvenating” star-formation history. However, previous works have highlighted the degeneracies between the SEDs of galaxies undergoing continuous and rejuvenating star-formation histories \cite{Witten2025-2}. A number of factors can affect this preferred SFH model, including emission line constraints and SFH parameters and priors. Nevertheless, both potential SFH models imply that YD4 has experienced significant, extended star formation in its past and is currently highly star forming, as a result of its large gas reservoir provided through tidal streams.}

The YD4's subcomponent YD6 shows a continuous star-formation history, with its SFR gradually increasing in the last $300\,{\rm Myr}$ and reaching a $10\,{\rm Myr}$ averaged SFR of $5.1\,{\rm M_{\odot}\,yr^{-1}}$ in the last time bin.
Although YD4 and YD6 may show different phase in their star-formation histories and they have large difference in their luminosities and morphology, they both steadily form stars as indicated by their sSFR of $6.3$ -- $13\,{\rm Gyr^{-1}}$.

YD7-E and ZD1 have a little \correct{recent} star-formation, while they have a more continuous star-formation histories in the past $10$ -- $150\,{\rm Myr}$, showing that these components are likely to be experiencing a quenching of star-formation in the last $10\,{\rm Myrs}$. These features are imprinted in their SEDs, that show only a weak emission line contributions to their photometry while they have prominent Balmer break features (Figure \ref{fig:SEDspectra}). 
In particular, YD7-E is luminous both in the rest-frame UV and optical showing relatively massive stellar mass of  $10^{8.9}\,{\rm M_{\odot}}$, and is characterized by its relatively blue UV spectral slope with a prominent Balmer break features, suggesting that YD7-E has experienced a recent quenching of star-formation activities.
The spectra of YD7-E indeed share a similar feature with JADES-GS-z7-QU \cite{Looser2024} which is  characterized as a temporarily quenched galaxy at a similar redshift.
These features are also consistent with the recent studies of quenched galaxies at high redshifts \cite{Covelo-Paz2025,Baker2025,Trussler2025}.

\section{\correct{Gravitational interactions in high-z multiple proto-galaxy systems in a cosmological simulation}}
Here we compare the observed properties of galaxy interactions using \correct{cosmological simulations of galaxy formation}. Although reproducing the exact properties of the \quintet\ using simulation is beyond the scope of this study, simulated galaxies can provide us with a theoretical background to understand physical properties we identified in the JWST and ALMA observations.
At the same time, it can provide us with a support for the interpretations we have made in the main text by exploring whether the observed phenomena are indeed expected in the gas-rich merger-induced galaxy evolution at high redshift.

To this end, we use a zoom-in cosmological simulation named {\it FirstLight} \cite{Ceverino2017}, which focuses on $z \gtrsim 6$ galaxies with a maximum resolving scale of $\sim17$ -- $32\,{\rm pc}$.
The cosmological box size is 60 comoving Mpc, within which all the galaxies with the circular velocities above $178\,{\rm km\, s^{-1}}$ at $z=5$ are targeted.
This corresponds to a stellar mass of $M_* \gtrsim 5\times 10^{9}\, M_\odot$ at $z=5$.
Compared to other high-resolution simulations \cite{Pallottini2019, Katz2022}, {\it FirstLight} provides a larger sample of massive galaxies at high redshift.
With  a parsec-scale resolution and a significant number of massive samples, {\it FirstLight} simulations are particularly suited to investigate the internal structure and evoutionary pathways of rare, gas-rich galaxy merger events such as the \quintet. 

\begin{figure} 
	\centering
	\includegraphics[width=0.8\textwidth]{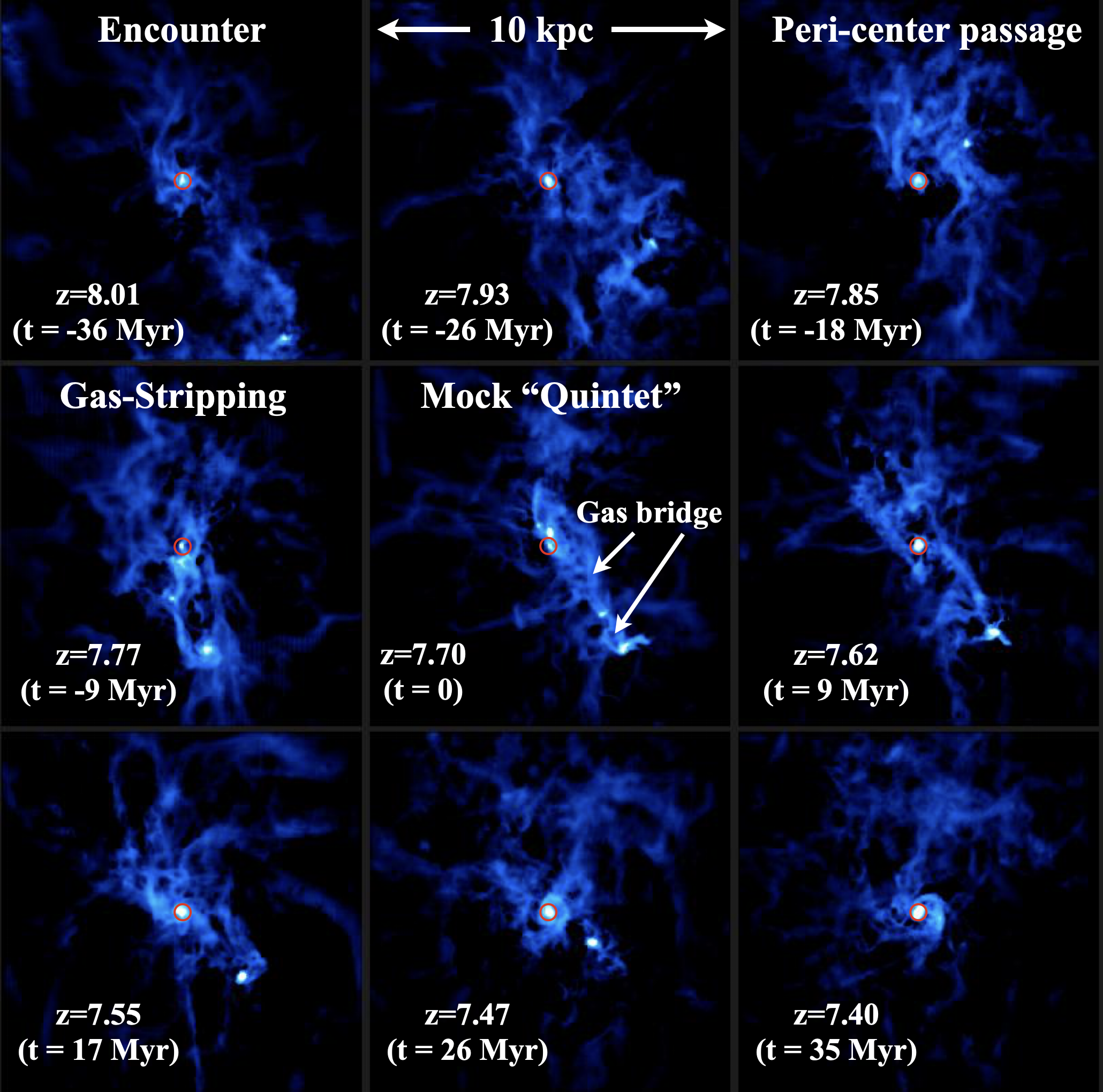}
	\caption{{\bf Time evolution of \correct{the surface density of cold  ($T<1000\,{\rm K}$)} gas distribution from $z=8.01$ to $z=7.40$ of simulated sample FL957 picked up from FirstLight simulations} The side and projected lengths are 10 kpc. The red circles represent the location of the \correct{center of the dark matter halo}. We set $t =0$ at $z=7.70$, where we identified at least four clumps that are luminous in \Oiiilong\ (see Figure \ref{fig:FL957_quintet} as well).
    }
	\label{fig:FL957_gas_projection} 
\end{figure}

We have successfully identified an analog of a high-redshift, gas-rich galaxy merger system, named FL957 at $z=7.7$ \correctn{(hereafter defined as $t=0$)}, whose properties have also been analyzed in previous studies \cite{Hashimoto2023, Nakazato2024}.
Figure \ref{fig:FL957_gas_projection} shows the time evolution of the projected gas density in this analog system, from $z=8.01$ \correctn{($t=-36\,{\rm Myr}$)} to $z=7.40$ \correctn{($t=+35\,{\rm Myr}$)}.
At $z=7.7$ \correctn{($t=0$)}, FL957 shows multiple gas and stellar clumps whose stellar masses are comparable to those observed in the \quintet\ ($\sim 4$ -- $7\times 10^8\, M_\odot$).
The maximum velocity offset among the simulated clumps is $131\, {\rm km\, s^{-1}}$, similar to the observed value of $193\pm 10\, {\rm km\,s^{-1}}$.
The comparable spatial size and kinematic scale of FL957 enable us to investigate process such as gas stripping, gas filamentary structure, and gas recycling that may produce galaxy components with diverse star formation histories and stellar ages.

In FL957, a galaxy major merger triggers the formation of multiple clumps, filamentary gas  strucures, and intense starbursts. The emergence of these features in the simulated system suggests that such processes can physically happen in the actual galaxies merger system like \quintet.
At $z=8.01$ \correctn{($t=-36\,{\rm Myr}$)}, an incoming galaxy begins interacting with a \correct{ galaxy, highlighted with a red circle in center of Figure \ref{fig:FL957_gas_projection}}, and undergoes a pericentric passage at $z=7.85$ \correctn{($t=-18\,{\rm Myr}$)}.
This interaction \correctn{temporarily} induces \correctn{a tidal removal} of gas from the \correct{halo center} located near the central galaxy (highlighted with a red circle in Figure \ref{fig:FL957_gas_projection}), also temporarily reducing its surface gas number density to $\sim 30\, {\rm cm^{-3}}$ \correctn{as calculated as an average using a $r=200\,{\rm pc}$} apeature. The corresponding free-fall time is about 10 Myrs, leading to temporary \correctn{decrease} of star formation of the central galaxy. 
The \correct{removed} gas is redistributed into extended tidal tails within which new clumps form.
These clumps give rise to new star formation without underlying older stellar population, resembling the situation of S1 and \correct{possibly YD1} in the \quintet.
Following this interaction, gaseous bridges emerge, connecting the clumps with typical  densities of $\sim 10\, {\rm cm^{-3}}$. 
The newly formed clumps begin to coalesce $\sim9$ -- $18\,{\rm Myrs}$ after their formation, further fueling star formation activity within these structures.
Nearly $50\,{\rm Myrs}$ after the second peri-center passage, these clumps were merged into a single massive galaxy by $z=7.40$ \correctn{($t=+35\,{\rm Myr}$)}, with a stellar mass of $6.3\times 10^9\,{\rm M_{\odot}}$, gas mass of $4.9 \times 10^9\,{\rm M_{\odot}}$, and a high star formation rate of ${\rm SFR}=63\,{\rm M_\odot\,yr^{-1}}$.
\correctn{With the elevaiting star formation rate soon after the interaction phase, the FL957 eventually become a massive ($M_{\ast}>10^{10}\,{\rm M_{odot}}$) galaxy at $z=6.6$ ($t\sim150\,{\rm Myr}$).}

\begin{figure} 
	\centering
	\includegraphics[width=0.8\textwidth]{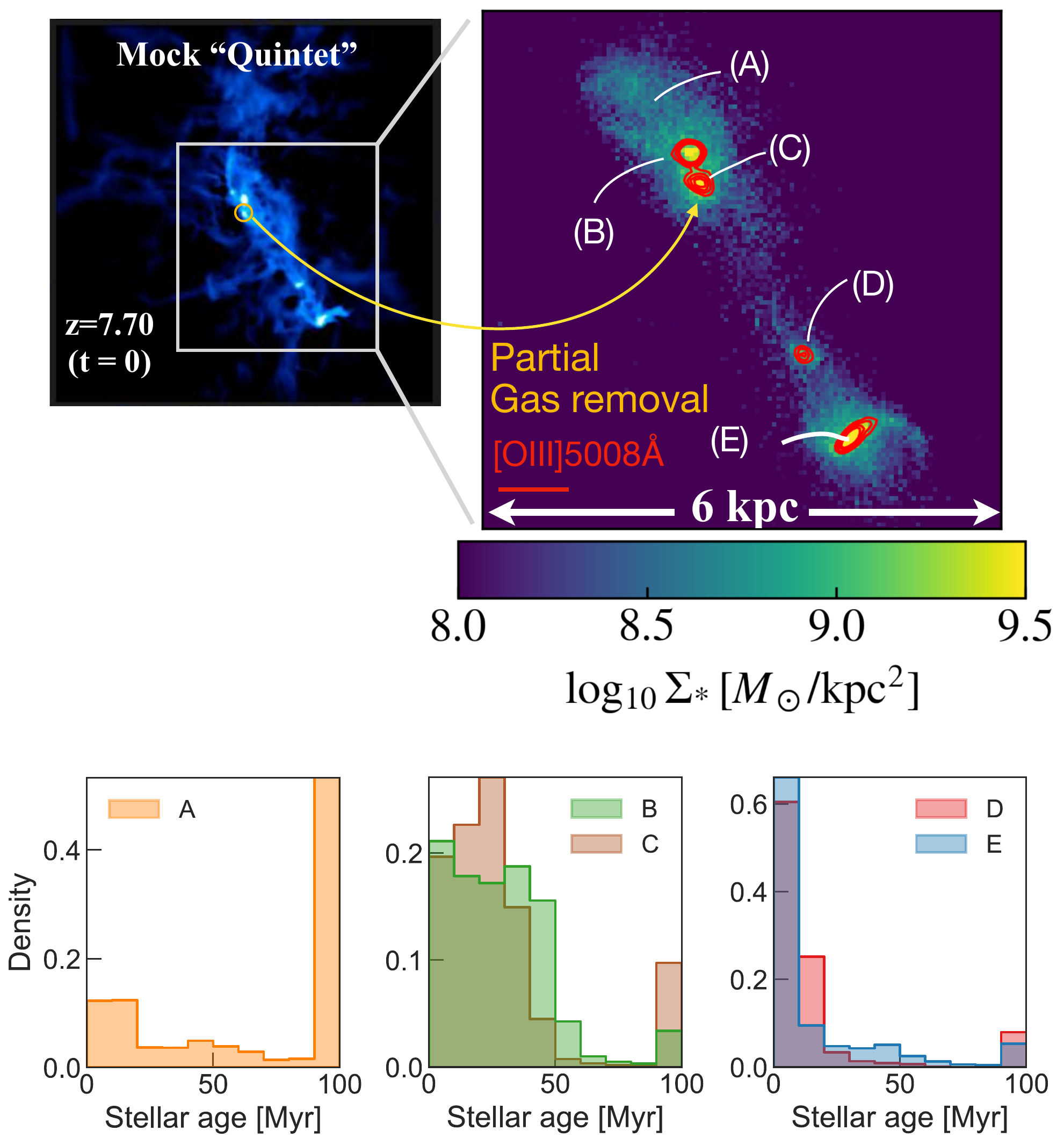}
	\caption{\textbf{Stellar population of galaxy clumps during a simulated merger event}:
	{\it Upper left panel}: The surface density of cold ($T<1000\,{\rm K}$) gas distribution of the FL957 at $z=7.70$.
	{\it Upper right panel}: The stellar mass surface density distribution of the merging galaxy and newly formed clumps at $z=7.70$. Red contours show \Oiiilong\ fluxes at levels of $2, 4, 6, 8, 10 \times 10^8 \, L_\odot\, {\rm kpc^{-2}}$. The yellow arrow indicates the position of the gas removed part (C) seen in the upper left panel, which have declining number of young stellar population in the mid-bottom panel.
	{\it Bottom panels}: Mass weighted stellar age distribution of part of merging galaxies marked from A to E in upper right, showing that matured (region A), more continuous (regions B and C), and young stellar (regions D and E) distribution exists in this system during a galaxy merger at high redshift.
	}
	\label{fig:FL957_quintet} 
\end{figure}

Figure \ref{fig:FL957_quintet} shows the stellar mass distribution during the most clumpy phase of FL957 at $z=7.7$. In this phase, different regions within the system show complex and diverse star formation histories, as shown by their mass-weighted stellar age distributions.
\correct{In the simulated galaxy merger in Figure \ref{fig:FL957_quintet}, the region A is} composed almost exclusively of \correct{mature} (age $\sim100\,{\rm Myrs}$) stellar populations and show little signs of on-going star formation, as indicated by the absence of emission lines such as  \Oiiiboth.
In contrast, other regions \correct{(regions D and E)} are dominated by young (age $\lesssim30\,{\rm Myrs}$) stellar populations with active star-formation. 
There are also transitional zones (regions B and C) where both old and young stellar populations coexist with gas-rich star-formation activities.
This diversity in stellar population parallels the observed properties of the \quintet, suggesting that such complexity in star formation history can naturally arise through gas-rich mergers at high redshift.

In summary, the cosmological simulation reproduces a morphology and properties similar to the observed \quintet\ system, demonstrating that the clumpy structure is formed through mergers, supporting the interpretations from the observed properties of the \quintet. The associated gas \correctn{removal} and subsequent star formation within the tidal tails contribute to the complex star formation histories observed in each component.

\correct{\section{Interpreting Galaxy Evolution in the \quintet\ through Combined ALMA and JWST Observations}}

\correct{We conclude that the evolution of the \quintet\ is primarily driven by  galaxy-galaxy interactions in the core of a dense protocluster environment.
Observations from ALMA reveal a complex distribution of neutral gas tracers, \Cii\ emission and dust continuum, including three bright \Cii\  clumps and two filamentary bridges connecting YD1, YD4, and YD7.
The morphology and alignment of these bridges are inconsistent with typical star-formation driven outflows, \correctn{which is expected to show bi-conical or symmetric, centrally diverging morphologies without a preferred alignment along intergalactic axes \cite{Walter2002}.}
Instead, they strongly indicate ongoing tidal gas exchange as a result of major mergers.}

\correct{Complementary SED fitting using JWST photometry reveals a wide range of star formation histories among the member galaxies. 
Some galaxies, such as YD1 and S1, show recent starbursts and YD4 shows  star formation activity over extended period, while others like ZD1 and YD7-E appear quiescent. 
This diversity aligns with the observed gas distribution: actively star-forming galaxies coincide with bright \Cii\, dust clumps, and gas-rich bridges, whereas quiescent components are largely devoid of neutral gas tracers.
These findings together with ALMA's results indicate that tidal interactions, stripping gas from some galaxies while fueling others, are modulating the star formation activity across the system.}

\correct{In light of the cosmological simulation results from the {\it FirstLight}, observed properties of the \quintet\ are consistent with theoretical expectations for gas-rich galaxy mergers at $z \gtrsim 7$. In particular, the analog system FL957 reproduces many of the observed characteristics: tidal gas stripping from interacting galaxies, the formation of clumpy in-situ star-forming regions within gaseous bridges, and diverse star formation histories across components.
The simulation also demonstrates that such interactions can cause both temporary quenching, through partial gas removal or redistribution, and the triggering of new star formation via accretion from tidal structures.
These processes provide a physical explanation for the combined ALMA and JWST observations of the \quintet, strengthening the interpretation that major mergers are shaping its current state.}

\correct{Taken together, the spatial distribution and kinematics of \Ciium-emitting, the dust continuum, the diversity in star formation histories, and the physical mechanisms reproduced in the simulation all support a unified interpretation: the galaxies in the \quintet\ are undergoing major mergers that induce both gas removal and gas inflow, driving the observed mixture of quenching, continuous star formation, and starburst episodes. This case study provides a rare, spatially resolved view of how dense early environments shape the evolutionary pathways of massive galaxies at the epoch of reionization.
}


\newpage

\newpage

\newpage

\newpage

\newpage
\begin{table}[h] 
	\centering
	\caption{\textbf{Dust continuum properties of the \quintet.} Continuum fluxes are measured value without gravitational lensing correction, while infrared luminosities and dust masses are lens-corrected by a factor of $2$. \correctn{A dust temperature of $T_d=50\,{\rm K}$, the dust emissivity index of $\beta_{\rm d}=1.5$, and the dust mass absorption coefficient of $\sim23.4\,{\rm cm^2\,g^{-1}}$ are applied.} \correct{The upper limits represent $3\,\sigma$ upper limits.}}
	\label{tab:dust} 
	\begin{tabular}{lccc} 
		\\
		\hline
        ID & $f_{\rm cont.}$ & Infrared Luminosity & Dust Mass\\
           & (${\rm \mu Jy}$) & $\times10^{10}\,{\rm L_{\odot}}$ & $\times10^{5}\,{\rm M_{\odot}}$\\
        \hline
        YD1 & $18\pm4$ & $3.0\pm0.7$ & $6.2\pm1.4$\\
        S1 & $<12$ & $<2.0$ & $<4.1$\\
        YD4 & $20\pm4$ & $3.3\pm0.7$ & $6.8\pm1.4$\\
        YD7-E & $<12$ & $<2.0$ & $<4.1$\\
        YD7-W & $15\pm4$ & $2.5\pm0.7$ & $5.1\pm1.4$ \\
        ZD1 & $<12$ & $<2.0$ & $<4.1$\\
		\hline
	\end{tabular}
\end{table}
\end{document}